\documentclass[traditabstract]{aa}
\usepackage{graphicx,float,psfrag}
\usepackage{tabularx}
\usepackage{latexsym,amsmath,amssymb}
\usepackage{natbib}
\usepackage{verbatim}
\usepackage{enumitem}

\bibpunct{(}{)}{;}{a}{}{,}

\def \george    {G. Younes}
\def \delphine  {D. Porquet}
\def \bassem    {B. Sabra}
\def \nicolas   {N. Grosso}
\def \mallen    {M.~G. Allen}
\def \jreeves   {J.~N. Reeves}

\def \strasbourg {Observatoire Astronomique de Strasbourg, UMR7550, Universit\'e de Strasbourg, CNRS, 11 rue de
l'Universit\'e, F-67000 Strasbourg, France}

\def \ndu {Department of Sciences,  Notre Dame University-Louaize, P.O.Box 72, Zouk Mikael, Lebanon}
\def \keele {Astrophysics Group, School of Physical \&\ Geographical Sciences, Keele University, Keele, Staffordshire ST5 5BG}

\newcommand {\hst} {\textsl{HST}}
\newcommand {\xmm} {\textsl{XMM-Newton}}
\newcommand {\chandra} {\textsl{Chandra}}

\def \rsun {\ifmmode$R$_{\odot}\else R$_{\odot}$}

\def \hcm {\hbox {\ifmmode $ atoms cm$^{-2}\else atoms cm$^{-2}$\fi}}

\def\approxgt{\mathrel{\hbox{\rlap{\lower.55ex \hbox {$\sim$}}
        \kern-.3em \raise.4ex \hbox{$>$}}}}
\def\approxlt{\mathrel{\hbox{\rlap{\lower.55ex \hbox {$\sim$}}
        \kern-.3em \raise.4ex \hbox{$<$}}}}

\def\arcsec{\hbox{$^{\prime\prime}$}}

\newcommand {\chisq} {$\chi ^{2}$}

\def \src {NGC~4278}

\begin{document}

\title{X-ray  and  multiwavelength  view  of \src.\\
A  LINER--Seyfert connection$?$}

\author{\george\inst{1} \and \delphine\inst{1} \and \bassem\inst{2} \and \nicolas\inst{1} \and \jreeves\inst{3} \and \mallen\inst{1}}

\institute{\strasbourg \and \ndu \and \keele}

%\date{Received; Accepted:}
\date{Received / Accepted}
%\date{Received / Accepted}

\authorrunning{G. Younes et al.}

\titlerunning{X--ray and multiwavlength view of \src}

\abstract{The emission  mechanism responsible  for the bulk  of energy
from radio to X--rays in low ionization emission line regions (LINERs)
and  Low  Luminosity Active  Galactic  Nuclei  (LLAGN)  has been  long
debated.  Based  on UV  to X--ray  and radio to  UV flux  ratios, some
argue  that  LINERs/LLAGN are  a  scaled-down  version  of their  more
luminous predecessors Seyfert galaxies.   Others, based on the lack of
X--ray short (hours) time--scale  variability, the non detection of an
iron line  at 6.4~keV, and the  faint UV emission  compared to typical
AGNs, suggest the  truncation of the classical thin  accretion disk in
the inner regions of the AGN where a radiatively inefficient accretion
flow  (RIAF)  structure   forms.  We  investigate  the  LINER--Seyfert
connection  by studying  the  unabsorbed LINER  galaxy  NGC 4278  that
accretes  at  a  low  rate  (L$_{bol/Edd}\sim$7$\times$10$^{-6}$)  but
exhibits a  broad H$\alpha$ line,  and a point-like nucleus  in radio,
optical,  UV  and  X-rays.  We   analyzed  one  XMM-Newton  and  seven
\chandra\  X-ray observations  of NGC  4278 spread  over a  three year
period,  allowing the  study of  the X--ray  variability  at different
time-scales  (hours, months, years).   We also  examined the  radio to
X-ray spectral energy distribution  to constrain the accretion mode in
the  nucleus  of  \src.    Long  time-scale  (months)  variability  is
observed  where  the  flux increased  by  a  factor  of $\sim$3  on  a
time-scale of a  few months and by a factor of  5 between the faintest
and the brightest observation  separated by $\sim$3 years.  During the
XMM-Newton observation,  where the highest flux level  is detected, we
found a  10$\%$ flux increase  on a short  time-scale of a  few hours,
while the light curves for the different \chandra\ observations do not
show short  time-scale (minutes to hours)  variability.  A combination
of    an   absorbed   power    law   ($N_{H}\approx10^{20}$~cm$^{-2}$,
$\Gamma=2.2^{+0.1}_{-0.2}$)      plus     a      thermal     component
(kT$\approx0.6$~keV) were able to fit the \chandra\ spectra. The \xmm\
spectra, where  the highest X--ray  flux is detected, are  well fitted
with an  absorbed power--law with no  need for a  thermal component as
the emission from the power--law component is dominant. The power--law
photon index  is $\sim2.1$ and the  hydrogen column density  is of the
order  of   $10^{20}$~cm$^{-2}$.   Neither   a  narrow  nor   a  broad
Fe~K${\alpha}$ emission line at 6.4~keV  are detected with a 22~eV and
118~eV  upper limits derived  on their  equivalent widths.   We derive
optical  fluxes  from archival  \hst\  ACS  observations and  detected
optical variability on time--scales of  years.  For the first time for
this  source, thanks  to  the  optical/UV monitor  on  board \xmm,  we
obtained simultaneous  UV and X-ray flux  measurements. We constructed
SEDs  based on  simultaneous  or quasi  simultaneous observations  and
compared them to LINER, radio--loud, and radio--quiet quasar SEDs.  We
find that at a low X--ray flux the \src\ SED resembles that of typical
LINER sources where the radio  to X--ray emission can be considered as
originating from  a jet  and/or RIAF, whereas  at a high  X--ray flux,
\src\ SED  is more like  a low luminosity Seyfert  SED.  Consequently,
\src\ could exhibit both  LINER and Seyfert nuclear activity depending
on the strength of its X--ray emission.}
%\abstract{T}{T}{T}


\maketitle

\section{Introduction}
\label{sec:intro}

Low-ionization  nuclear  emission  line  regions (LINERs)  were  first
identified by \citet{heckman80aap} as a class of galaxies with optical
spectra dominated  by emission lines from low  ionization species. The
ionization  mechanism is  yet  poorly known  and  could be  explained,
either           in            terms           of           starbursts
\citep{alonso-herrero00ApJ:starburstinliners}  or,  more consistently,
being  due  to  a   low  luminosity  active  galactic  nuclei  (LLAGN)
\citep{terashima00ApJ:liners,  ho93ApJ:linerAGN}.   This  latter  idea
rises  from  the  detection  of  broad  H$\alpha$\  components  in  an
important  fraction   of  LINER  sources   \citep[noted  as  LINER~1.9
(hereinafter LINERI) objects][]{ho97apjs},  and/or a point--like UV or
X-ray  source at  the  nucleus.   If the  luminosity  scales with  the
accretion  rate, the  study  of  the LINER  nucleus  define a  supreme
pattern  for probing  low accretion  rate physics  around supermassive
black holes (SMBHs).

How similar are these LINERI sources to classical luminous Seyfert and
quasar galaxies?   Based on observational properties,  the weakness or
absence   of   a   big   blue   bump   feature   at   UV   wavelengths
\citep{ho08aa:review}       usually        detected       in       AGN
\citep{malkan82apj:uvexcSey1,sanders1989ApJ:bbbseyqua,koratkar99pasp:seybbb},
the   lack  of  a   broad  Fe~K$\alpha$   emission  line   at  6.4~keV
\citep{terashima02apjs:LLAGNASCA,ptak04apj:ngc3998},  except  for  the
peculiar LINER  NGC~1052 \citep{brenneman09ApJ:ngc1052}, and  last but
not  least, the  lack of  short time-scale  (minutes to  hours) X--ray
variability  have  been  attributed  to  an  intrinsic  difference  in
LINERsI/LLAGN central  engine as  opposed to normal  Seyfert galaxies.
One  proposed   scenario  is   that  accretion  in   LINERsI/LLAGN  is
radiatively inefficient,  advection dominated compared  to the typical
geometrically thin optically thick accretion disks present in luminous
AGN \citep{narayan05apss:adaf}.

However,   more  recently  \citet{maoz05apj:linervarUV}   revealed  UV
variability in the  nuclei of a sample of 17  LINERs observed with the
\hst.   By combining  these results  with non--simultaneous  radio and
X--ray   observations,  \citet{maoz07MNRAS}   demonstrated   that  the
UV/X--ray luminosity  ratios are similar to those  of Seyfert~1 nuclei
and pointed  out that LLAGN may  be a scaled--down  version of Seyfert
galaxies where a  thin accretion disk exists.  This  idea is supported
by   \citet{pianmnras10}  who  studied   simultaneous  UV   to  X--ray
observations of four  LINER nuclei observed with the  XRT and the UVOT
on--board the  \textsl{Swift} telescope.  They  discovered short--time
scale  (half a day)  X--ray variability  in two  of their  objects and
showed that the UV to X--ray  flux ratios are consistent with those of
more luminous AGN.

The elliptical  galaxy \src\ \citep[distance of  16.7~Mpc;][ scaled to
$H_{0}=70$~km~s$^{-1}$~Mpc$^{-1}$]{tonry01apj:dist}  has  been studied
extensively at different wavelengths.   At radio wavelength, a compact
non-thermal   radio  source  has   been  detected   at  6   and  18~cm
\citep{jones84apj:4278radobs}.      \citet{4278nagar05aap}    reported
two-sided radio emission on subparsec  scales in the form of twin jets
emerging from  a central compact  component ($T_{B}=1.5\times10^9$~K).
\citet{ho97apjs}  classified  \src\  as  a  type 1.9  LINER  from  the
definite    detection    of    a    broad   H$\alpha$    line,    with
$log~F(H\alpha)=-13.07$~ergs~cm$^{-2}$~s$^{-1}$,  supporting  the  AGN
nature of the nuclear engine.  Moreover, an unresolved compact nuclear
source     has    been    detected     in    \hst\     WFPC2    images
\citep{capetti00aa:hst}.  Finally,  \citet{ho01apjl}, after studying a
\chandra\ 1.4 ks snapshot taken in  April 2000, gave \src\ a class (I)
X-ray     morphology,     showing  a  dominant     nuclear     source.
\citet{terashima03apj:rloud}  fit  the  0.5--8 keV  spectrum  obtained
during  this same  \chandra\ snapshot  with  a power  law modified  by
absorption   and   found   a   2--10~keV   corrected   luminosity   of
$9.1\times10^{39}$~ergs~s$^{-1}$.   \citet{gonzalezmartin09aa}  find a
0.5--10~keV  corrected  luminosity of  $5.6\times10^{39}$~erg~s$^{-1}$
after  studying a  $\sim$100~ks \chandra\  observation taken  in March
2006.

The mass  of a  black hole in  the nucleus  of \src, derived  from the
M-$\sigma$        relation        \citep{termaine02ApJ:Mbh},        is
$3.09\pm0.54\times10^{8}$~M$_{\odot}$
\citep{wang03MNRAS:Mbh,chiaberge05ApJ:Mbh}.     This   leads    to   a
$L_{Edd}\approx3.9\times10^{46}$~ergs~s$^{-1}$;      assuming     that
$L_{Edd}=1.25\times10^{38}~(M_{BH}/M_{\odot})$~ergs~s$^{-1}$.      This
implies a very low L$_{X}/$L$_{Edd}$ ratio of 2$\times10^{-7}$.

In this  paper we  report a  timing and X--ray  spectral study  of the
LINER galaxy \src\  observed with \xmm\ and \chandra.   We describe in
Sect.~2  the X-ray observations  and the  data reduction.   Timing and
spectral X--ray results as well as optical (\hst/ACS) and UV (\xmm/OM)
results  are  presented  in  Sect.~3.   In Sect.~4,  we  describe  the
construction of the \src\ spectral energy distribution.  We discuss in
sect.~5 our  results in the context of  the LINER--Seyfert connection.
The main results are summarized in Sect.~6.

\section{X-ray observations and data reduction}
\label{sec:reduction}

%-----------
% Table  1
%-----------
\begin{table*}[!th]
\label{obs-param}
\caption{Log of the X-ray observations.}
\newcommand\T{\rule{0pt}{2.6ex}}
\newcommand\B{\rule[-1.2ex]{0pt}{0pt}}
\begin{center}{
\begin{tabular}{c c c c c}
\hline
\hline
\textbf{Satellite} \T \B & \textbf{Instrument} & \textbf{Start date} & \textbf{Obs. ID} & \textbf{Exposure time (ks)} \\
\hline
\chandra \T & ACIS-S & 2000 April 20 & 398 & 1.4 \\
\xmm$^a$ \T & EPIC & 2004 May 23 & 205010101 & 34/35.5/35.5 \\
\chandra \T & ACIS-S & 2005 February 02 & 4741 & 37 \\
\chandra \T & ACIS-S & 2006 March 16 & 7077 & 110 \\
\chandra \T & ACIS-S & 2006 July 25 & 7078 & 51 \\
\chandra \T & ACIS-S & 2006 October 24 & 7079 & 105 \\
\chandra \T & ACIS-S & 2007 February 20 & 7081 & 110 \\
\chandra \T & ACIS-S & 2007 April 20 & 7080 & 55 \\
\hline
\end{tabular}}
\end{center}
\begin{list}{}{}
\item[Note:]$^{a}$Exposure time for pn/MOS1/MOS2 respectively.
\end{list}
\end{table*}
%-----------
% Table  1
%-----------

\subsection{\chandra\ observations}
\label{chan-obs}

\src\ was observed seven times  with the \chandra\ observatory, with a
first  1.4~ks snapshot taken  in 2000  April 20  \citep{ho01apjl}. The
following  six  observations  were  taken  over a  two  year  interval
starting on  2005 February 03 and  ended on 2007 April  20.  The seven
observations    were   obtained    with   the    spectroscopic   array
\citep[ACIS-S;][]{weisskopf02PASP} where the nucleus was placed on the
aim point  of the ACIS-S3  back-illuminated chip.  They were  taken in
Very  Faint   mode  to  increase   their  sensitivity.   All   of  the
observations    are   \chandra\    archival    data   obtained    from
chaser~\footnote{\label{chaser}http://cda.harvard.edu/chaser/Descriptions.}.
The log of the \chandra\ observations are listed in Table~1.

All  seven observations  were reduced  and analyzed  in  a systematic,
homogeneous way using the CIAO software package version 4.2, \chandra\
Calibration Database,  CALDB, version 4.2,  and the ACIS  Extract (AE)
software package version  3.175 \footnote{\label{AEnote} The {\em ACIS
Extract}  software   package  and   User's  Guide  are   available  at
http://www.astro.psu.edu/xray/acis/acis\_analysis.html.}
\citep{broos2010AE}.   We  started  by  using  the level  1  event  file
produced by the \chandra\ X-ray  Center (CXC) to suppress the position
randomization  applied  by  the  CXC  Standard  Data  Processing  when
creating a level  2 event file.  We also corrected  for the effects of
charge-transfer  inefficiency  on   event  energies  and  grades.   We
filtered for bad event  grades (only ASCA grades 0, 2, 3,  4 and 6 are
accepted) and hot columns to  take account of several potential issues
such as cosmic  rays and bad pixels. Good  time intervals, supplied by
the pipeline, were applied to the final products
\footnote{\label{ciao1note} Description on how to create a new level 2
event          file          can          be         found          at
http://cxc.harvard.edu/ciao/threads/createL2/}.

In order to get the X-ray position for the sources in the ACIS-S field
of  view,  we  used  a  wavelet  transform  detection  algorithm,  the
\textit{wavdetect}  program  within  the  CIAO  data  analysis  system
\citep{wavdetect:freeman02apjs}.   The  output  file of  the  previous
procedure is  input to  the AE  software.  The AE  package is  used to
refine  source  positions,  extract  source photons,  construct  local
backgrounds,  extract  source,  and  background spectra  (see  below),
compute  redistribution  matrix files  (RMFs)  and auxiliary  response
files  (ARFs), by spawning  the \textit{mkarf}  and \textit{mkacisrmf}
routines of CIAO, and perform spectral grouping and fitting.  We focus
here  on  the  X-ray  emission  from  the  nuclear  source  \citep[for
information    on    the   X-ray    source    population   of    \src\
see][]{brassington09apjs}.

Source  events are  extracted  around the  source  centroid, inside  a
polygonal     shape    of    the     local    PSF,     generated    by
MARX\footnote{http://space.mit.edu/ASC/MARX/} at  the energy 1.497 keV
using  the \textit{ae\_make\_psf} tool  implemented in  AE. Background
region was defined on the brightest of the seven observations (obs ID:
4741) following  the  AE  procedure.   This background  region  is  an
annular region centered on the  source position where the inner radius
delimit 1.1~$\times$~99\% encircled energy radius and the outer radius
is set such  that the background includes 200  counts.  This number is
obtained   from  a  special   image  where   all  events   within  the
$\sim1.1\times99$\% PSF circles  of all the sources in  the field were
excluded  (\textsl{swiss cheese  image}).  This  background  region is
then generated to the other observations in order to have a consistent
and  coherent measurement  of  the diffuse/unresolved~X--ray  emission
between the seven observations for a homogeneous spectral analysis.

We  find that  all of  the  seven \chandra\  observations suffer  from
pile-up.  X-ray data affected by pile-up can mimic spectral and timing
variability.  Therefore,  pile-up was  accounted for by  excluding the
core of the PSF (see Appendix~A  for more details).  We decided not to
use the first 1.4~ks  \chandra\ snapshot observation of \src.  Indeed,
after removing the  piled-core, not enough counts remained  to build a
useful spectrum.  We use the tool \textit{dmextract}, called by the AE
software,  to   create  a  source  spectrum  over   the  energy  range
0.5--8~keV.  We used the tool \textit{ae\_group\_spectrum} implemented
in  AE to  group the  spectra.   Channels between  0.5 and  8 keV  are
grouped to have a three sigma ($3\sigma$) signal to noise ratio, which
corresponds to  a minimum of 20 counts  per bin, to enable  the use of
$\chi^{2}$ statistics in the spectral analysis.

\subsection{\xmm\ observation}

The  log of the  \xmm\ observation  is listed  in Table~1.   \src\ was
observed by \xmm\ on 2004 May 23.  The EPIC-pn \citep{struder01aa} and
MOS \citep{turner01aa}  cameras were  operated in Imaging,  Prime Full
Frame  mode  using the  thin  filter  and  medium filter  respectively
(exposure time  about 34 and  35.5 ks, respectively).   The Reflection
Grating Spectrometers show few counts therefore they were not included
in   our   analysis.    The   optical/UV   monitor   (OM)   instrument
\citep{mason01aa:om} was  operated in imaging  mode with the  UVW1 and
the  UVM2   filters  (effective   $\lambda$  of  2910   and  2310~\AA,
respectively).  All  data products  were obtained from  the XMM-Newton
Science                                                        Archive
(XSA)\footnote{http://xmm.esac.esa.int/xsa/index.shtml}   and  reduced
using the  Science Analysis System  (SAS) version 9.0.  The  EPIC data
were  not affected  by pile-up,  and no  severe intervals  of enhanced
solar activity  were present.  Data are selected  using event patterns
0--4 and  0--12 for pn and  MOS, respectively, during  only good X-ray
events (``FLAG$=$0'').

We extract source events from  a circle of $10\arcsec$ radius centered
on the  source.  Background events  are extracted from  a source--free
circle of the same radius on the same CCD as the source.  We generated
response  matrix  files  using  the SAS  task  \textit{rmfgen},  while
ancillary   response  files   were  generated   using  the   SAS  task
\textit{arfgen}.  The  EPIC spectra were  created in the  energy range
0.5--8~keV  to  enable   flux  and  model--parameter  comparison  with
\chandra\   spectra\footnote{In  fact,   the  contribution   from  the
surrounding medium around \src\ that we calculated using the \chandra\
images is  unknown below  0.5~keV as we  extracted the spectra  in the
0.5--8~keV  range,  see sect.~3.3.2  for  more  details.}.  They  were
grouped using the FTOOLS \textit{grppha}  tool to have a minimum of 20
counts per  bin to allow the  use of the $\chi^2$  statistic.  The SAS
task \textit{omichain} was used to process the OM data.

\section{Results of the nucleus study}

\subsection{Lightcurves and hardness ratios}
\label{lightcurveSec}

\begin{figure}[]
\centerline{\includegraphics[angle=0,width=0.5\textwidth]{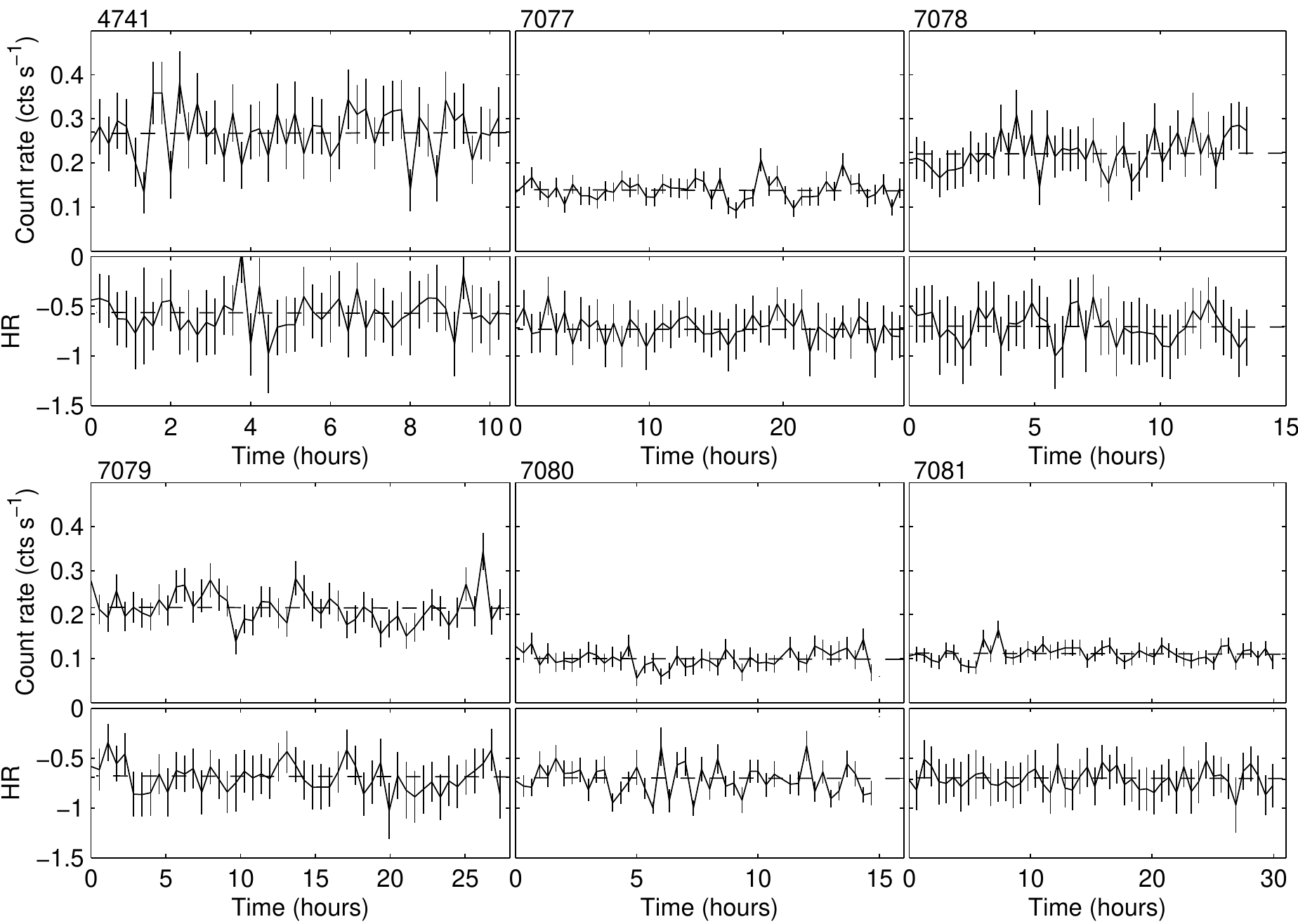}}
\caption{Light curves of the six \chandra\ observations along with their hardness ratios defined as  $HR=(H-S)/(H+S)$, where $S$ is the count rate in the soft 0.5-2 keV band and $H$ is the count rate in the hard 2-10 keV band. Light curves were binned to have 50 groups with a signal to noise ratio greater than 10. The dashed lines show the averages of the count rates and hardness ratios.}
\label{light-hardness}
\end{figure}

Light curves for the six  \chandra\ observations were extracted in the
energy range  0.5-10 keV from  the pile-up free  wings of the  PSF. To
enable  the comparison  between  the different  observations, the  net
count rate of each is then  corrected for the excluded fraction of the
PSF. The hardness ratio is  defined as, $HR=(H-S)/(H+S)$, where $S$ is
the count rate in the soft 0.5-2 keV band and $H$ is the count rate in
the hard 2-10 keV band. Fig. \ref{light-hardness} shows the 0.5-10 keV
light curves and the hardness-ratios of the different observations. No
clear-cut short  timescale, minutes to hours,  variability is detected
in any of the  observations. We conducted a Kolmogorov--Smirnov, K--S,
test to check more accurately any variability in the source count rate
within each observation.  We find  a K--S test probability $>17$\%\ in
the  six observations  that  the nuclear  emission  originates from  a
constant source.  Therefore, \src\  is considered as constant on short
(minutes  to hours)  timescale.   On the  other  hand, long  timescale
(months) variability  is observed between  the different observations.
The hardness  ratio in all of  the six observations  indicate that the
source is in  a soft state being constant  during each observation and
between   the   observations  as   well   implying  similar   spectral
characteristics.

We extracted an EPIC-pn light  curve from a 10\arcsec\ circle centered
on       the        source.        We       used        the       tool
\textit{epiclccorr}\footnote{http://xmm.esa.int/sas/current/doc/epiclccorr/epiclccorr.html}
to  correct  for  good  time  intervals (GTIs),  dead  time,  and  for
background subtraction.  On account  of the low angular resolution of
\xmm\  compared to  \chandra, events  from the  10\arcsec\  circle are
contaminated by emission from point--like sources and diffuse emission
(Fig. \ref{xmm-vs-chandra}). Nevertheless, we  show in Appendix B that
emission from the  central source is dominant and  all the surrounding
sources  only contribute  in 2\%\  of  the total  X-ray flux.   Figure
\ref{pn-light-hardness} shows  the pn  light curve and  the associated
hardness  ratio.  According to  a K--S  test, \src\  shows variability
during the  \xmm\ observation with  a K--S probability of  0.8\%\ that
the emission comes  from a constant source. This  variability is clear
at the beginning  of the observation where we  detect a 10\%\ increase
over a period of $\sim$1~hour.

\begin{figure}[]
\centerline{\includegraphics[angle=0,width=0.5\textwidth]{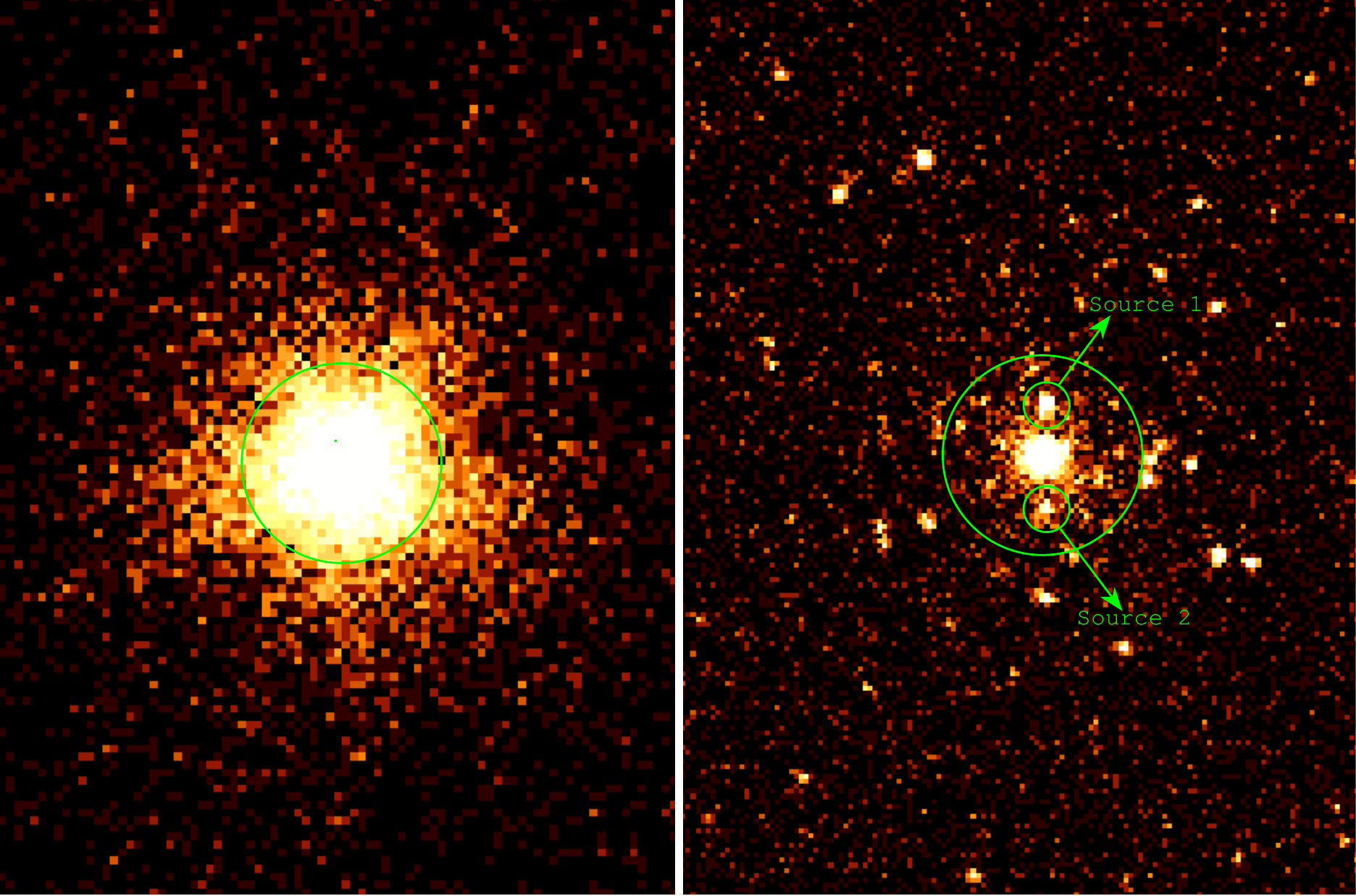}}
\caption{Left panel: the \xmm/MOS1 image of \src\ with a 10\arcsec\ circle centered on the nucleus. Right panel: the \chandra\ image (obs ID: 7077) with a 10\arcsec-radius circle centered on the nucleus. The lower angular resolution of \xmm\ has blended the central source, X-ray point--like sources, and diffuse emission, detected with \chandra, in one point--like source.}
\label{xmm-vs-chandra}
\end{figure}

\begin{figure}[]
\centerline{\includegraphics[angle=0,width=0.5\textwidth]{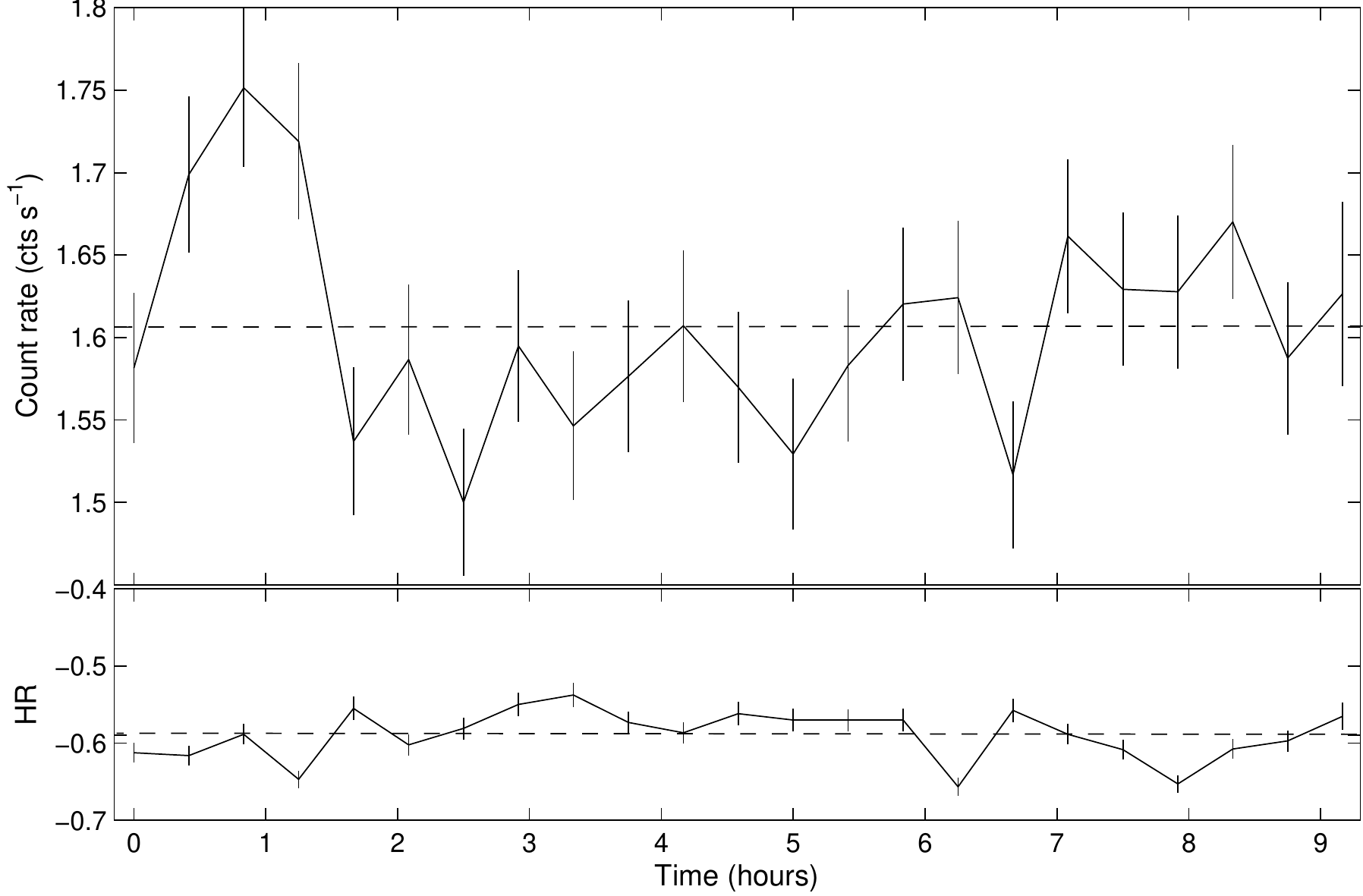}}
\caption{Upper panel: EPIC-pn light curve of \src\ during the \xmm\ observation with a time bin size of 1.5~ks. Lower panel: hardness ratio. The dashed lines show the average on the count rate and hardness ratio.}
\label{pn-light-hardness}
\end{figure}

Finally,   we   calculated   the   normalized   ``excess   variance'',
$\sigma^{2}_{rms}$  defined  as  in  \citet{nandra97apj:variance}  and
\citet{turner99apj:seyvar}  for the  seven light  curves, in  order to
check   more   accurately  any   possible   variability  during   each
observation.   We  find that  $\sigma^{2}_{rms}<0.01$  in  all of  the
observations with  an error $<15$\%  and $\sim$20\% for  the \chandra\
and \xmm\  observations simultaneously.   This result is  discussed in
sect.~5.2.

\subsection{Spectral analysis}

The spectral  analysis was performed  using XSPEC \citep{arnaud96conf}
version  12.5.1. The  updated  photo-electric cross  sections and  the
revised solar abundances of  \citet{wilms00ApJ} are used throughout to
account for  absorption by neutral  gas.  An absorbed  galactic column
density derived  from \citet{kalberla05aa:nh} (obtained  with the W3NH
tool\footnote{http://heasarc.gsfc.nasa.gov/cgi-bin/Tools/w3nh/w3nh.pl})
and  fixed to  $2.07\times10^{20}$~cm$^{-2}$  was applied  to all  the
models.   Spectral uncertainties  are given  using  $\Delta$\chisq\ of
2.71, corresponding to 90\%  confidence for one interesting parameter,
and to 95\% confidence for upper limits.

\subsubsection{Chandra spectral analysis}
\label{SpecAna}

\begin{figure}[]
\centerline{\includegraphics[angle=0,width=0.5\textwidth]{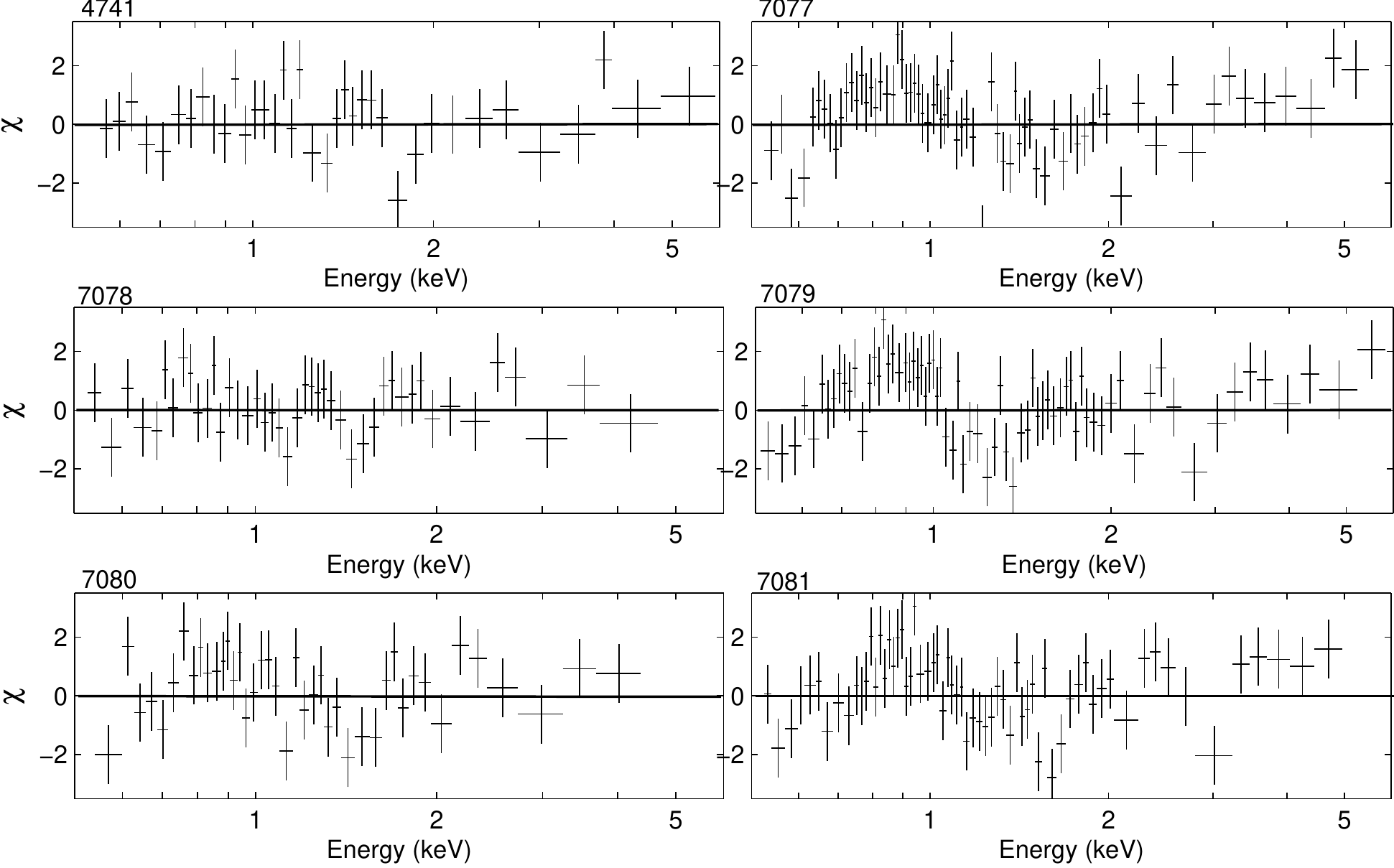}}
\caption{The residuals in terms of sigma of an absorbed power-law fit to the different \chandra\ observations. Some features are clearly present in the soft 0.5-2 keV band suggesting emission from diffuse hot gas.}
\label{resid-pow}
\end{figure}

\begin{figure}[!th]
\centerline{\includegraphics[angle=0,width=0.5\textwidth]{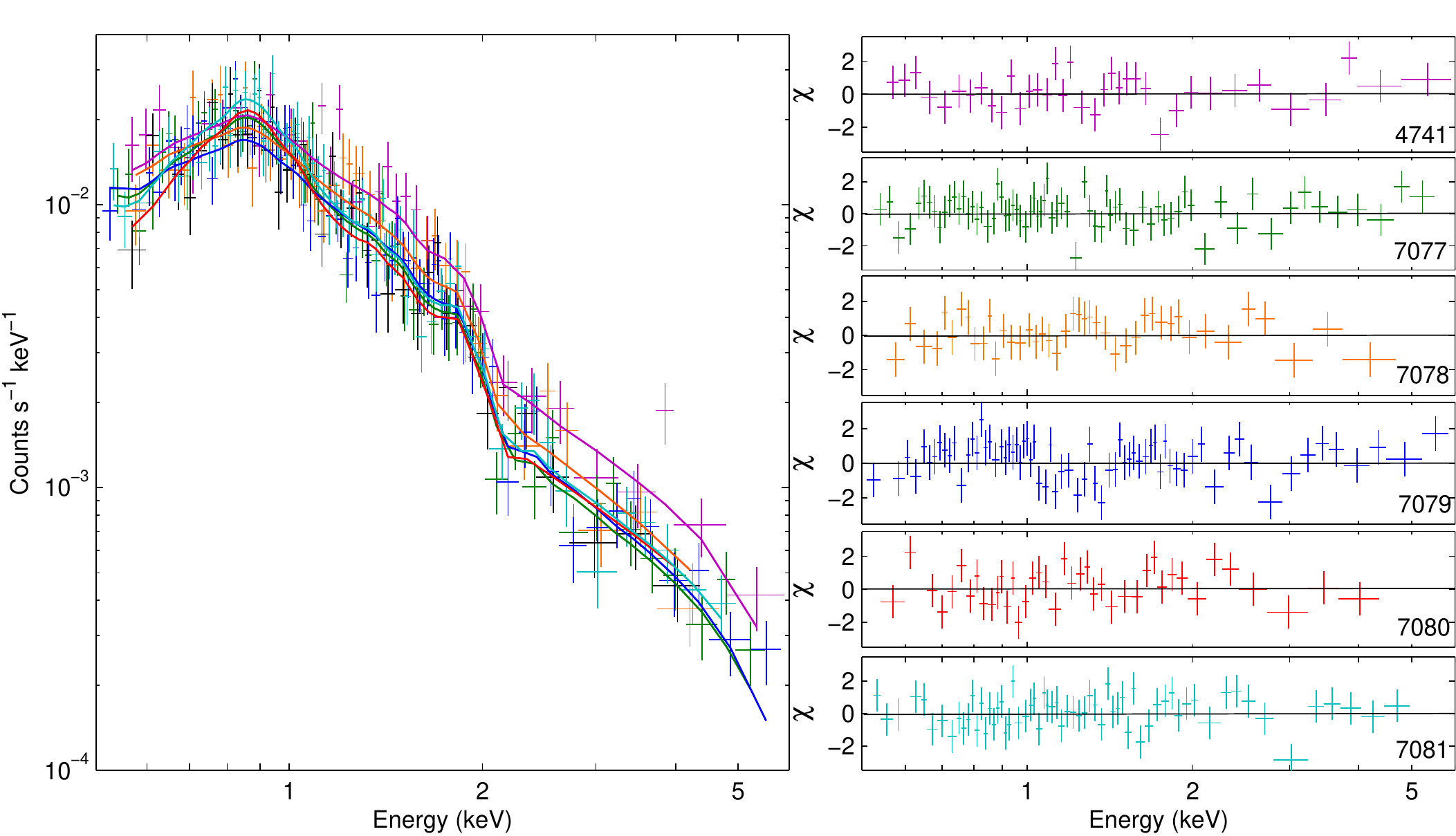}}
\caption{Left panel: data and best--fit simultaneous model of an absorbed power-law plus \textit{mekal} to the six \chandra\ observations. Right panels: the residuals of the fit in terms of sigma. The features in the soft band were accounted for by the thermal component. See the electronic edition of the Journal for a colour version of this figure.}
\label{pow-mek-fig}
\end{figure}

\begin{table*}[!th]
\label{specfit-TZTZPM}
\caption{Best fit parameters of a \textit{mekal} plus an absorbed power-law models for the \chandra\ data and an absorbed power--law model for the \xmm\ spectra.}
\newcommand\T{\rule{0pt}{2.6ex}}
\newcommand\B{\rule[-1.2ex]{0pt}{0pt}}
\begin{center}{
\resizebox{\textwidth}{!}{
\begin{tabular}{l c c c c c c c c}
\hline
\hline
Obs. ID \T \B & Nh$^{a}$ &$\Gamma$ & Pl norm. & kT$^{a}$ & EM$^{ab}$ & Flux$^{c}$ & Corr. flux$^{c}$ & Pl$^{d}$\\
 & $10^{20}$~cm$^{-2}$ & & $10^{-4}$~Photons~keV$^{-1}$~cm$^{-2}$~s$^{-1}$ &  keV & 10$^{62}$~cm$^{-3}$ & $10^{-13}$~erg~s$^{-1}$ & $10^{-13}$~erg~s$^{-1}$ & \%\ \\ 
\hline
 4741\T & $<$6.78 & 2.13~[2.00-2.28] & 4.28~[3.92-4.71] & 0.62~[0.58-0.66]  & 2.64~[2.27-2.99] & 18.1~[17.1-18.8] & 18.5~[18.2-20.1] & 94\\
 7077\T & & 2.26~[2.16-2.39] & 1.82~[1.68-1.99] &  &  & 7.8~[7.5-7.9] & 8.21~[8.0-9.0] & 83\\
 7078\T & & 2.34~[2.22-2.47] & 4.21~[3.90-4.61] &  &  & 16.0~[14.9-16.2] & 16.5~[16.1-18.5] & 93\\
 7079\T & & 2.38~[2.28-2.50] & 3.76~[3.54-4.07] &  &  & 14.1~[13.3-14.3] & 15.05~[14.7-16.4] & 90\\
 7080\T \B & & 2.02~[1.84-2.20] & 1.10~[0.96-1.26] &  & & 5.9~[5.5-6.1] & 6.0~[5.7-6.8] & 80\\
 7081\T \B & & 2.12~[2.00-2.25] & 1.25~[1.14-1.29] &  & & 6.2~[5.9-6.4] & 6.4~[6.2-7.1] & 81\\
\hline
\multicolumn{9}{c}{Reduced $\chi^{2} = 0.93$ for 310 d.o.f.} \T \B \\
\hline
 0205010101 \T \B & 3.82~[3.05-4.59] & 2.05~[2.03-2.07]  & 8.07~[7.90-8.24] & & & 33.8~[32.8-33.2]& 34.5~[34.2-34.8] & 100 \\
\hline
\multicolumn{9}{c}{Reduced $\chi^{2} = 1.01$ for 487 d.o.f.} \T \B \\
\hline
\end{tabular}}}
\end{center}
\begin{list}{}{}
\item[Note:]$^{a}$Parameter linked for the different \chandra\ observations.
\item[]$^{b}$The emission measure (EM) of the \textit{mekal} model, EM=$\int$n$_{e}$n$_{H}$dV.
\item[]$^{c}$Fluxes in the 0.5--8~keV energy range.
\item[]$^{d}$The power-law component fraction to the total corrected flux.
\end{list}
\end{table*}

To obtain a  first guess about the models that  would be best suitable
to  the  data,  we  first   fit  the  six  pile-up  corrected  spectra
separately. An absorbed power-law model gives an acceptable fit in the
six cases  with a  power-law photon index,  $\Gamma$, between  2.3 and
2.8.  The  intrinsic hydrogen  column density  has an  upper  limit of
$1.2\times10^{21}$~cm$^{-2}$.   In five of  the six  observations, the
soft 0.5-2 keV band  shows residuals around $\sim$1~keV suggesting the
possible    presence    of    emission    from   hot    diffuse    gas
\citep[e.g.,][]{flohic06apj}.     Figure~\ref{resid-pow}   shows   the
residuals from  the single absorbed  power-law fit.  We note  that the
only observation that  did not show any evidence  for residuals around
$\sim$1 keV  is the  one with  the least net  source counts  (Obs. ID:
4741).   The  addition of  a  thermal component  \citep[\textit{mekal}
model;][]{mewe85aaps:mekal}  takes account of  these residuals  in the
soft band and improves the fit quality in the six observations.

Next, we fit  the six spectra simultaneously in  order to identify the
best  physical model  suitable for  the six  \chandra\ spectra  and to
better  constrain the model  parameters.  For  this purpose,  we tried
three different models:

\begin{enumerate}[]
\item Two power-laws affected by absorption, where the first power-law
represents the  soft 0.5--2~keV emission  from the AGN and  the second
power--law represents the  2--10~keV compton up--scattered emission by
a corona.

\item  A power-law  affected  by intrinsic  absorption  and a  thermal
component.   The power-law  represents the  direct  continuum emission
from the AGN whereas the thermal component takes into account possible
emission from unresolved X-ray binaries and/or supernovae remnants.

\item A power--law and a  thermal component affected by absorption and
an  additional power-law  component  \citep{guainazzi05aa:sey2}.  This
model  is  similar  to  model  (2) but  includes  the  possibility  of
Compton--scattered X-ray emission.
\end{enumerate}

Model (1)  gives an acceptable fit  with a reduced  $\chi^{2}$ of 1.38
for  311 d.o.f.   However, residuals  around  1 keV  persisted in  the
different  spectra.  We  then tried  model (2)  leaving  the intrinsic
absorption component, the power-law, and the \textit{mekal} parameters
free for  the six spectra.   The temperature and the  emission measure
(EM) of the thermal component, as well as the hydrogen column density,
were   consistent  within   the  error   bars  during   the  different
observations and therefore linked between each other.  The fit gives a
reduced $\chi^{2}$  of 0.93  for 310 d.o.f.   The mean average  of the
photon  index, $\Gamma$, is  2.2$_{-0.2}^{+0.1}$.  The  temperature of
the  thermal component is  $0.62\pm0.04$~keV.  The  intrinsic hydrogen
column  density  affecting  the  power--law  has  an  upper  limit  of
$6.78\times10^{20}$~cm$^{-2}$. According to  an F-test the statistical
improvement  in  the   fit  between  model  (1)  and   (2)  is  highly
significant. Finally,  we tried model (3)  used to fit  the spectra of
Compton-thin Seyfert~2 galaxies. Similar  to model (2), the parameters
of the thermal  component and the hydrogen column  density were linked
for the  different spectra. We find  a reduced $\chi^{2}$  of 0.93 for
309 d.o.f. The  absorption hydrogen column density has  an upper limit
of $8\times10^{20}$~cm$^{-2}$.  The average value of the photon index,
$\Gamma$, is $2.3\pm0.2$.  The temperature of the thermal component is
$0.63_{-0.05}^{+0.04}$~keV.   The addition of  a second  power-law did
not improve  the quality  of the fit.   The parameters of  the thermal
model and  the power--law model did  not change between  model (2) and
model  (3) while  the hydrogen  column density  is  better constrained
using model (2).  Moreover, according to an F-test, the improvement of
the fit by adding a power-law component is not significant.

Therefore, we consider model (2), an absorbed power--law and a thermal
\textit{mekal}  component,  as best  suitable  for  the six  \chandra\
spectra and we  show in Fig.  \ref{pow-mek-fig} the  data and the best
fit model  with the residuals in  the right panels. Table  2 gives the
parameters of  the best fit  model, the uncorrected and  the corrected
0.5--8~keV  flux, as  well as  the  power--law fraction  to the  total
corrected flux.

\subsubsection{\xmm\ spectral analysis}
\label{xmmSpecAna}

The \chandra\  image of \src\  (obs. ID: 7077)  is shown in  the right
panel of  Fig.  \ref{xmm-vs-chandra}.  The central  bright source, the
\src\ nucleus,  is surrounded by a 10\arcsec-radius  circle.  The left
panel is the  \xmm/MOS1 image with a circle of  the same size centered
on  the source.   The lower  angular resolution  of \xmm\  compared to
\chandra\ blended  the resolved and unresolved point  sources, as well
as the diffuse emission, in one point source.  In order to exclude the
contribution from  the contaminating sources,  the six spectra  of the
two brightest objects,  marked source 1 and source  2 in the \chandra\
image, obtained  in the energy range 0.5--8~keV  from each observation
were  fit simultaneously  with  an absorbed  power-law, assuming  that
these    sources    are    Low    Mass    X-ray    Binaries    (LMXBs)
\citep{kim04ApJlmxb}.   We  then  excluded  the two  sources  and  the
central  object  and  fit  the  rest, obtained  in  the  energy  range
0.5--8~keV, with a combination of  a thermal component and a power-law
assuming  contribution  from  diffuse  emission and  unresolved  point
sources.   The point  sources and  the  diffuse emission  do not  show
significant  variability between  the six  \chandra\  observations and
therefore we assumed them to  be constant (detailed results on the two
point--like sources and the diffuse emission are presented in Appendix
B).   When fitting  the  \xmm\ spectra  obtained  from the  10\arcsec\
circle  around the  point--like source,  we included  the contribution
from all  of the above  (Source~1, source~2, and the  diffuse emission
that contribute  only to 2\%\  of the total  flux, see appendix  B for
more details), the rest is then attributed to the nucleus.

\begin{figure}[!th]
\centerline{\includegraphics[angle=0,width=0.5\textwidth]{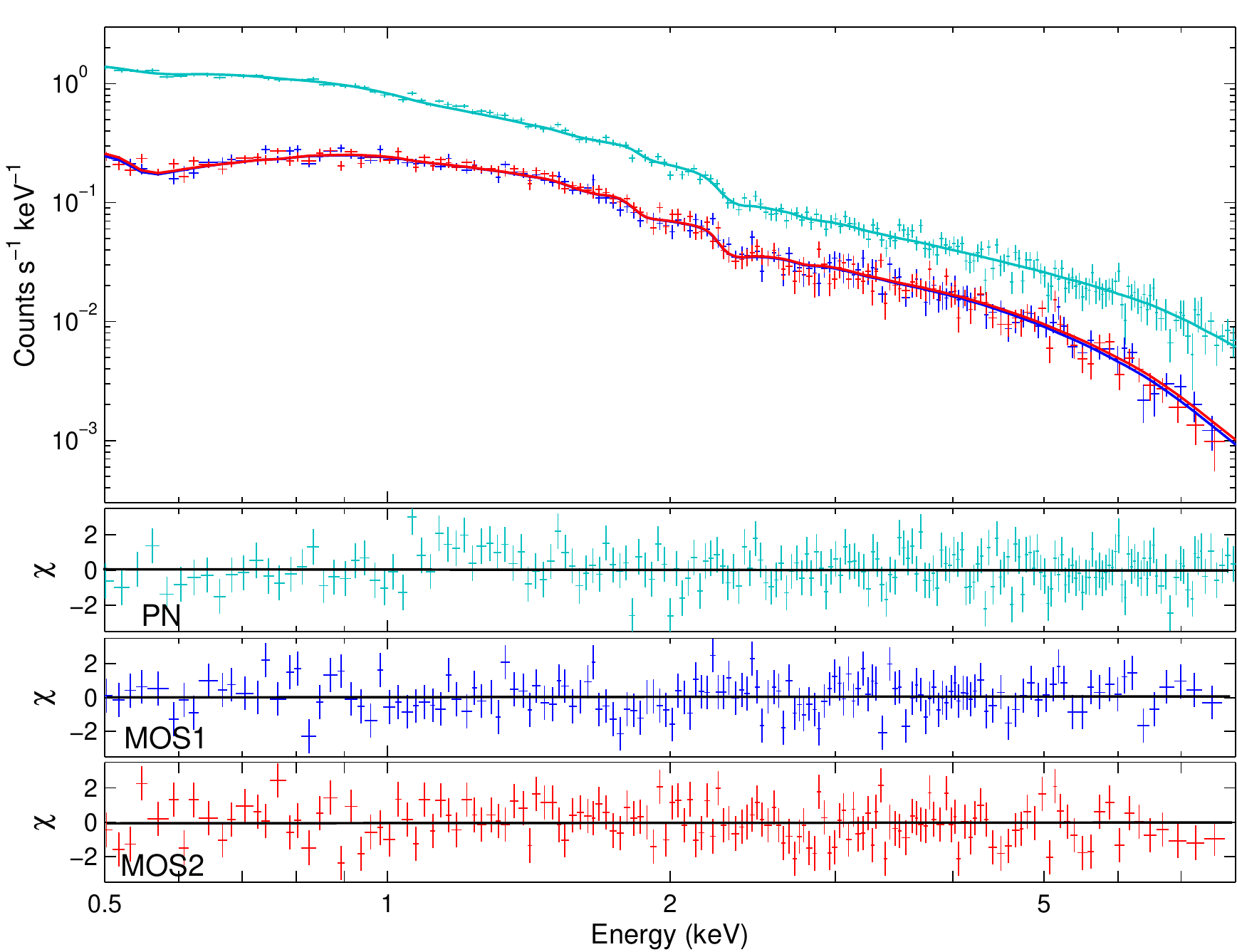}}
\caption{Upper panel: data and best fit model of an absorbed power-law to the PN, MOS1 and MOS2 (from top to bottom). The three other panels show the residuals of the fit in terms of sigma.}
\label{pow-mek-fig-xmm}
\end{figure}

We  started the analysis  by fitting  the pn,  MOS1, and  MOS2 spectra
simultaneously   with  an   absorbed  power-law   model   leaving  the
normalization  free  to  take  account for  slight  cross  calibration
uncertainties  between the  three  detectors.  The  fit is  noticeably
acceptable  with a  reduced $\chi^{2}$  of  1.01 for  487 d.o.f.   The
photon index,  $\Gamma$, and the  normalization of the  power--law are
$2.05\pm0.02$                                                       and
$(7.73\pm0.17)\times10^{-4}$~photons~keV$^{-1}$~cm$^{-2}$~s$^{-1}$
respectively.     The   intrinsic    hydrogen   column    density   is
$3.82\pm0.77\times10^{20}$~cm$^{-2}$.   We  note  that this  value  is
better  constrained than  the value  derived from  the  \chandra\ fit.
Then, we tried  to fit the \xmm\ spectra with the  same models used on
the \chandra\ data. The combination of two absorbed power--laws, model
(1), did  not give  a better fit  than one single  absorbed power--law
with a reduced  $\chi^{2}$ of 1.02 for 484 d.o.f.   The addition of an
unabsorbed  thermal component  to the  absorbed  power--law, model(2),
gives  an acceptable fit  with a  reduced $\chi^{2}$  of 0.99  for 482
d.o.f.   An  F-test  indicates   that  the  probability  for  such  an
improvement (compared to the fit  with a single absorbed power law) to
occur by chance  is 0.25.  Moreover, the emission  from the power--law
component dominates the  total nuclear emission to a  99\%\ level even
at low energies.  Therefore, the  thermal component is not required by
the  data  and  was not  kept  in  the  model.   Model (3),  where  an
additional  absorbed  power--law  is  added  to  model  (2),  did  not
ameliorate the  quality of the fit  with a reduced  $\chi^{2}$ of 1.01
for 479  d.o.f.  Therefore, the  absorbed power--law model  is favored
for the  \xmm\ spectra.  We give  in Table~2 the  best fit parameters.
Fig.  \ref{pow-mek-fig-xmm} shows the  data, folded model, and the fit
residuals.

No evidence is  found for an Fe~K${\alpha}$ emission  line at 6.4~keV,
neither in  the pn spectrum nor  in the \chandra\  spectra.  The upper
limit derived on its equivalent  width (EW) is measured from the \xmm\
observation as the  spectral resolution of the EPIC-pn  is higher than
that  of the  ACIS-S \chandra\  instrument.  By  including  a Gaussian
profile with a width fixed to 0  and 0.3~keV (for a narrow and a broad
line  respectively) at 6.4~keV  we derive  upper--limits of  22~eV and
118~eV on the  EW's of the narrow and broad  components to the 6.4~keV
line  respectively, which  are  commonly detected  in  the spectra  of
normal                           Seyferts                          and
quasars\citep{porquetaa04:21pgquasar,jiminez05AA:pgquasarfeline}.

%See the electronic edition of the  Journal for a color version of this
%figure.

\subsection{Optical data}

\subsubsection{\xmm\ optical/UV monitor}

The optical/UV monitor (OM) detector  on board of \xmm\ observed \src\
for six  2880~s exposures in the  UVM2 and UVW1  filters.  The central
region of \src\ in the UVM2 filter is detected as a point--like source
not showing  any sign of diffuse  emission from the  host galaxy.  The
SAS  tool \textit{omichain}  produced an  OM  source list  based on  a
3\arcsec\ aperture  and background taken  from an aperture  with radii
3.7\arcsec--4.3\arcsec.  The nuclear corrected  count rate in the UVM2
is  0.56~counts~s$^{-1}$  which  corresponds  to  a  flux  density  of
$1.18\times10^{-15}$~ergs~cm$^{-2}$~s$^{-1}$~\AA$^{-1}$ (using a count
rate         to        flux        conversion         factor        of
$2.103\times10^{-15}$~ergs~cm$^{-2}$~\AA$^{-1}$~counts$^{-1}$
\footnote{http://web.archive.org/web/20050322143642/
xmm.vilspa.esa.es/sas/documentation/watchout/uvflux.shtml}).
\citet{ngc4278:cardullo09} show the observation of \src\ made with the
\hst\ WFPC2/F218 filter (band--pass: 2000--2500\AA, comparable to UVM2
band--pass: 2050--2450\AA) in 1994 and  in 1995.  It is clear in their
figures  that  the  dominant  source  of  emission  in  this  specific
band--pass is the central source.  Therefore, we can consider that the
flux  calculated  from the  UVM2  filter  corresponds  to the  nuclear
flux. We decided not to use  the UVW1 image as it shows strong diffuse
emission, most likely from the host galaxy, around the central source.

\subsubsection{\hst}

\src\  was observed  four times  with each  of the  ACS/WFC  F475W and
F850LP  filters between  December 2006  and January  2007.   For every
filter, we  derive the  flux for two  pointings where the  source lies
away from the CCD edge. We use the flux derived from the Hubble Legacy
Archive    (HLA)\footnote{http://hla.stsci.edu/hlaview.html}   DAOPhot
tool,   originally    in   counts~s$^{-1}$,   and    convert   it   to
ergs~cm$^{-2}$~s$^{-1}$     using      the     conversion     equation
$F_{\lambda}=C/(\varepsilon_{f}\int$S$_{\lambda}d\lambda)$  where C is
the count rate, $\varepsilon_{f}$ is  the fraction of the point source
energy encircled  within N pixels,  and $\int$S$_{\lambda}d\lambda$ is
the total  imaging point source  sensitivity\footnote{more details can
be  found at http://www.stsci.edu/hst/acs/documents/handbooks/cycle18/
c09\_expcalc3.html}.   We   find  a  flux  in  the   F475W  filter  of
$6.64\times10^{-13}$~ergs~cm$^{-2}$~s$^{-1}$                        and
$8.02\times10^{-13}$~ergs~cm$^{-2}$~s$^{-1}$ for  observations made in
December 23 2006 and January  2 2007 respectively.  The fluxes derived
for the  F850LP filter are $7.45\times10^{-13}$ergs~cm$^{-2}$~s$^{-1}$
and $8.08\times10^{-13}$ergs~cm$^{-2}$~s$^{-1}$  for observations made
on December 23 2006 and January 2 2007 respectively.

\section{Spectral energy distribution}

\begin{figure}[]
\centerline{\includegraphics[angle=0,width=0.5\textwidth]{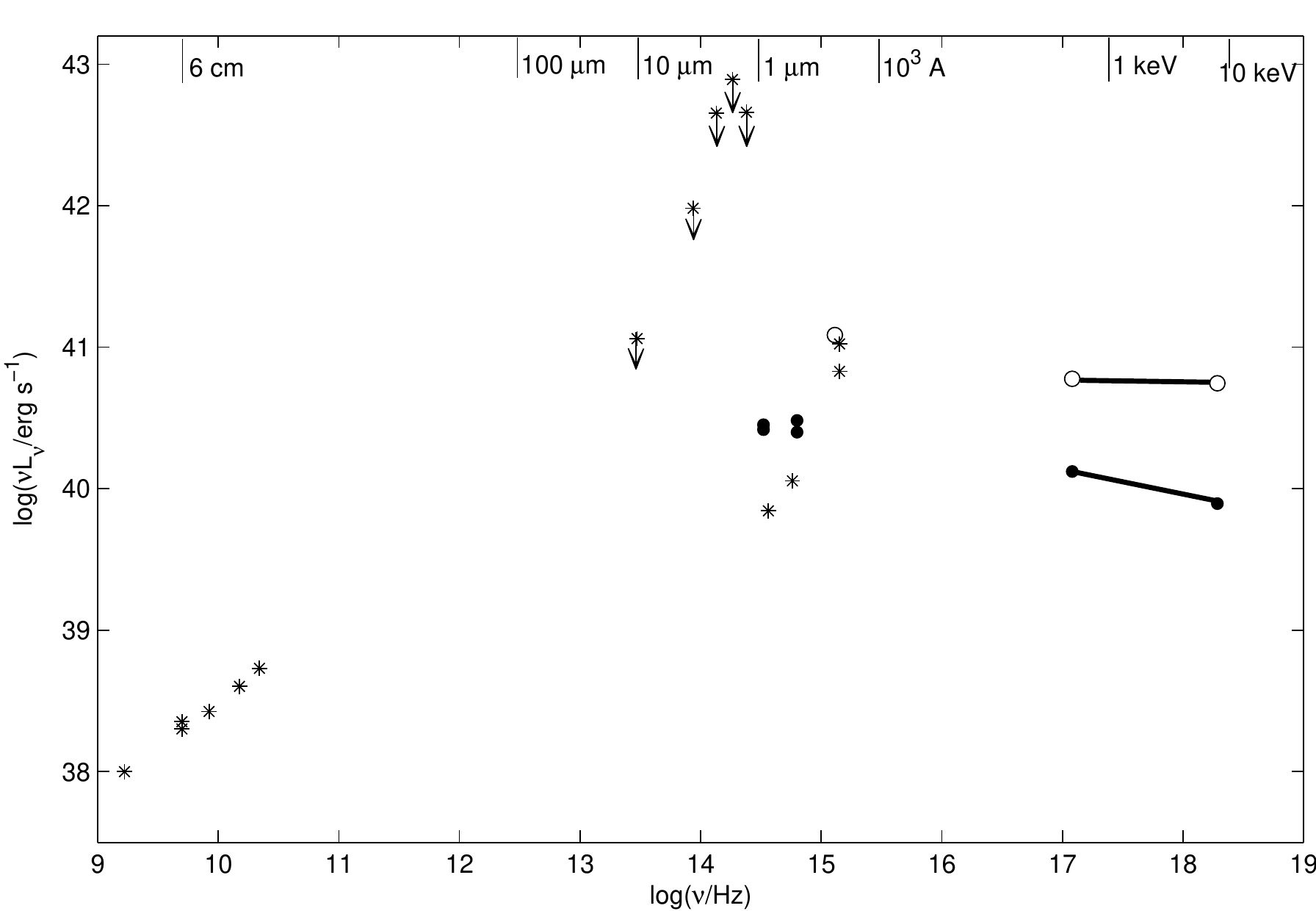}}
\caption{The spectral energy distribution of \src. Stars derive from non simultaneous radio to UV observations. The open circles are the simultaneous \xmm\ UV and X--ray fluxes. The dots represent the contemporary \hst/ACS optical and the \chandra\ (obs. ID: 7081) X--ray fluxes.}
\label{sed-fig}
\end{figure}

We compile the available photometry  data of the nucleus of \src\ from
the literature and add the data  derived in this work to construct the
\src\  SED. In  order  to  minimize any  contamination  from the  host
galaxy,  high  angular  resolution  is required  at  all  wavelengths.
Therefore, only data coming from observations with apertures less than
10\arcsec\ are used.  Table 3 lists  the data used to derive the \src\
SED shown in Fig. \ref{sed-fig}.

%-----------
% Table  3
%-----------
\begin{table*}[!th]
\label{sed-tab}
\caption{Multiwavelength nucleus data for \src.}
\newcommand\T{\rule{0pt}{2.6ex}}
\newcommand\B{\rule[-1.2ex]{0pt}{0pt}}
\begin{center}{
\begin{tabular}{l c c c c c}
\hline
\hline
\textbf{$\nu$} \T \B & \textbf{$\nu L_{\nu}$} & Aperture& Satellite/filter & Date & Reference\\
(Hz)                 & (ergs~s$^{-1}$)        &  (arcsec)        &                  &      & \\
\hline
1.93$\times10^{18}$  &    7.83$\times10^{39}$ &   3.34           & \chandra/ACIS-S & February 2007  & This work    \\
1.93$\times10^{18}$  &    5.56$\times10^{40}$ &   10             & \xmm/EPIC-pn & May 2004  & This work        \\
1.21$\times10^{17}$  &    1.32$\times10^{40}$ &   3.34           & \chandra/ACIS-S & February 2007  & This work   \\
1.21$\times10^{17}$  &    5.96$\times10^{40}$ &   10             & \xmm/EPIC-pn  & May 2004 & This work        \\
1.42$\times10^{15}$  &    1.05$\times10^{41}$ &   0.28           & \hst/F218W & January 1995 &\citet{ngc4278:cardullo09}\\
1.42$\times10^{15}$  &    6.75$\times10^{40}$ &   0.28           & \hst/F218W & June 1994 & \citet{ngc4278:cardullo09}\\
1.30$\times10^{15}$  &    1.21$\times10^{41}$ &   3              & \xmm/UVM2 & May 2004  & This work \\
6.32$\times10^{14}$  &    3.03$\times10^{40}$ &   0.15           & \hst/F475W & January  2007 & This work \\
6.32$\times10^{14}$  &    2.51$\times10^{40}$ &   0.15           & \hst/F475W & December 2006 & This work \\
5.76$\times10^{14}$  &    1.13$\times10^{40}$ &   1.26           & \hst/F555W & May 1994 & \citet{lauer05aj:hst}     \\
3.65$\times10^{14}$  &    6.95$\times10^{39}$ &   $<$6           & \hst/F814W & May 1994 & \citet{capetti02aap}      \\
3.32$\times10^{14}$  &    2.83$\times10^{40}$ &   0.15           & \hst/F850LP & January  2007 & This work \\
3.32$\times10^{14}$  &    2.61$\times10^{40}$ &   0.15           & \hst/F850LP & December 2006 & This work \\
2.40$\times10^{14}$  &    4.56$\times10^{42}$ &   8.00           & 1.54~m Mount Lemmon/J &  & \citet{heckman83apj:irobs}\\
1.82$\times10^{14}$  &    7.77$\times10^{42}$ &   7.20           & 3.8~m UKRIT/H &  & \citet{longmore82mnras:irobs}\\
1.36$\times10^{14}$  &    4.50$\times10^{42}$ &   7.20           & 3.8~m UKRIT/K &  & \citet{longmore82mnras:irobs}\\
8.69$\times10^{13}$  &    9.57$\times10^{41}$ &   8.00           & 1.54~m Mount Lemmon/L &  & \citet{heckman83apj:irobs}\\
2.97$\times10^{13}$  &    1.15$\times10^{41}$ &   5.60           & 1.54~m Mount Lemmon/N &  & \citet{heckman83apj:irobs}\\
2.20$\times10^{10}$  &    5.36$\times10^{38}$ &   $<$1           & VLA & August 2003 & \citet{giroletti05apj:ngc4278}\\
1.50$\times10^{10}$  &    4.00$\times10^{38}$ &   0.15           & VLA & September 1999 & \citet{nagar01:radobs} \\
8.40$\times10^{9}$   &    2.66$\times10^{38}$ &   (3.2$\times$1.9)$\times10^{-3}$& VLBA/VLA & August 2000 &  \citet{giroletti05apj:ngc4278}\\
5.00$\times10^{9}$   &    2.00$\times10^{38}$ &   (3.2$\times$1.9)$\times10^{-3}$& VLBA & July 1995 &   \citet{giroletti05apj:ngc4278}\\
5.00$\times10^{9}$   &    2.25$\times10^{38}$ &   2.5$\times10^{-3}$& VLBA/VLA & August 2000 &  \citet{giroletti05apj:ngc4278}\\
1.67$\times10^{9}$   &    1.00$\times10^{38}$ &   5.0$\times10^{-3}$& VLBI & May 1979 &  \citet{jones84apj:4278radobs} \\
\hline
\end{tabular}}
\end{center}
\begin{list}{}{}
\item[Note:]Infrared, optical and UV luminosities are corrected for galactic extinction.
\end{list}
\end{table*}
%-----------
% Table  3
%-----------

In radio wavelengths, and in  order to minimize the contamination from
the  radio  jet emission,  milli-arcsec  data,  measured through  VLBI
techniques, are used when  available. Otherwise, VLA observations with
sub-arcsec resolution  are applied.  All of  the near-IR (1--3~$\mu$m)
data for \src\ should be  considered as upper limits as the resolution
of those  observations is between  5\arcsec\ and 10\arcsec.   Only the
10-20~$\mu$m  mid-IR point  could be  considered as  less  affected by
emission  from normal stellar  populations and  may represent  to some
extent the nuclear emission (thermal  emission from dust grain can not
be ruled  out).  Optical  and UV data  are obtained from  Hubble Space
Telescope  (\hst)   observations  where  $~1$\arcsec\   apertures  are
used. One  UV data point  coming from the  \xmm\ OM telescope  is used
(see sect.   3.3.1). X-ray data,  measured in this paper,  and derived
from \chandra\ and \xmm\ observations are used.

The    Galactic    extinction     toward    \src\    is    0.11    mag
\citep{devaucouleurs91}.  The same value is derived using the Galactic
hydrogen    column     density    and    applying     the    equations
E(B-V)$=$NH/(5.8$\times10^{21}$)~mag     and     A$_{V}$/E(B-V)$=$3.1.
Internal extinction is reported  in \citet{ho97apjs} to be 0 according
to   a    Balmer   decrement   H$\alpha$/H$\beta$=2.50    leading   to
E(B-V)=0. Infrared,  optical and UV data are  de-reddened according to
these  extinctions  and  adopting  the Galactic  extinction  curve  of
\citet{cardelli89apj:gal.ext}.

The  radio   slope  between  5   and  22  GHz  is   somewhat  inverted
($\alpha\sim-0.6$), typical  for LINERsI/LLAGN \citep{ho99sed}.  Radio
loudness  is quantified  using the  radio  to X-ray  ratio defined  as
$R_{X}=\nu L_{\nu}(6cm)/L_{X}$, where $L_{X}$ is the luminosity in the
2-10 keV band \citep{terashima03apj:rloud}.  Following this criterion,
\src\  is  considered as  a  radio--loud  LINER  with $R_{X}$  ranging
between     $2.4\times10^{-2}$     and    $3.5\times10^{-3}$     where
$R_{X}=3.162\times10^{-5}$  is the  boundary for  radio  loud systems.
There is  a tendency for  a maximum in  the SED at  mid-IR wavelengths
where stellar population emission  is less effective, although thermal
emission from dust grains can not  be ruled out. The UVM2 flux derived
from the  OM telescope is  used to derive  an optical to  X--ray slope
$\alpha_{ox}$   and  a   radio  loudness   parameter   $R_{UV}$  where
$\alpha_{ox}$=0.384~log~($L_{2keV}$/$L_{2500\AA}$)                  and
$R_{UV}$=log($L_{\nu}(6cm)/L_{\nu}(2500)$\AA).   These  two parameters
are later used  for comparison purposes with LINER  and normal Seyfert
galaxies.      We      derive     a     2~keV      luminosity     $\nu
L_{\nu}(2~keV)\approx4\times10^{40}$~erg~s$^{-1}$  from  the  unfolded
model of the \xmm\ pn spectrum which gives an $\alpha_{ox}$ of $-1.2$.
We emphasize that  this value of $\alpha_{ox}$ is  determined, for the
first  time for  \src, from  simultaneous UV  and  X-ray observations.
$R_{UV}$  is about  2.7 and  the 2-10~keV,  $\alpha_{x}$ ($\Gamma-1$),
spectral index is $\sim$1.2.

\section{Discussion}

\subsection{X--ray properties}

The \src\ \chandra\ observations are well fitted with a combination of
an absorbed power--law and a thermal component. The temperature of the
thermal model is $0.62\pm0.04$~keV for the different observations. The
power--law    photon   index,    $\Gamma$,   has    an    average   of
$2.2_{-0.2}^{+0.1}$.   As  for  the  \xmm\  observation,  an  absorbed
power--law alone is sufficient to fit the spectra well with a $\Gamma$
of  $\sim2.1$.  The  intrinsic hydrogen  column density  affecting the
power--law is of the order  of $10^{20}$~cm$^{-2}$ in both cases.  The
fluxes   derived   for  the   \chandra\   observations  vary   between
$5.9\times10^{-13}$  and  $18.1\times10^{-13}$~ergs~cm$^{-2}$~s$^{-1}$
while   the    flux   measured   for   the    \xmm\   observation   is
$33.8\times10^{-13}$~ergs~cm$^{-2}$~s$^{-1}$. \citet{terashima03apj:rloud}
studied a sample of LLAGN nuclei, that display a compact radio source,
observed with \chandra.  The  authors found that the 0.5-8~keV spectra
of all the sources could be explained with an absorbed power--law with
most of the objects having a  hard X--ray spectrum with a photon index
$<$2,  including  \src.   Our  analysis  of the  \chandra\  and  \xmm\
observations  showed  a   somewhat  softer  0.5--8~keV  spectrum  with
$\Gamma$   greater  than   2.   The   \src\  observation   treated  by
\citet{terashima03apj:rloud}  was  affected  by  pile--up at  a  level
greater than  10\%\ and at  such a high  level, the pile--up  model of
\citet{davis01apjpile},   used   by  \citet{terashima03apj:rloud}   to
correct  for   pile--up,  is  misleading.   \citet{gonzalezmartin09aa}
studied a sample of 82 LINERs observed with \chandra\ and/or \xmm\ and
fit  the  spectra,  in  most  of  the cases,  with  a  combination  of
power--law and a  thermal component.  The authors find  a median value
for  the  photon  index,  $\Gamma$,  of $2.11\pm0.52$  and  a  thermal
component temperature of $0.54\pm0.30$~keV, compatible with the values
we find  for \src.   The authors find  a bimodal distribution  for the
2--10~keV  corrected  luminosities   they  derived  for  their  sample
centered       at       L(2--10~keV)$=10^{39}$~ergs~s$^{-1}$       and
L(2--10~keV)$=10^{41}$~ergs~s$^{-1}$. At  a lower flux,  \src\ lies in
the low  luminosity part  of the group  with a 2--8~keV  luminosity of
$8\times10^{39}$~ergs~s$^{-1}$  while at higher  flux, \src\  moves to
the   high   luminosity   group   with  a   2--8~keV   luminosity   of
$6\times10^{40}$~ergs~s$^{-1}$.                           Additionally,
\citet{gonzalezmartin09aa}  studied  one  \chandra\ observation  (obs.
ID: 7077) of \src\ and found a photon index of $2.6_{-0.3}^{+0.1}$ and
a  thermal temperature  of $0.5^{+0.2}_{-0.1}$~keV  comparable  to the
values  we derive  for the  same observation.   However, they  find an
observed               0.5--10~keV               flux               of
$\sim$4$\times10^{-13}$~ergs~s$^{-1}$~cm$^{-2}$  which   is  half  the
value    we    derive   for    the    same   observation.     However,
\citet{gonzalezmartin09aa} did not correct for the $\sim$8\%\ pile--up
fraction that  could explain the  discrepancy between the  two derived
flux values as pile--up  tempt to underestimate flux measurements (see
Appendix A).

\subsection{Variability}

\begin{figure}[!t]
\centerline{\includegraphics[angle=0,width=0.5\textwidth]{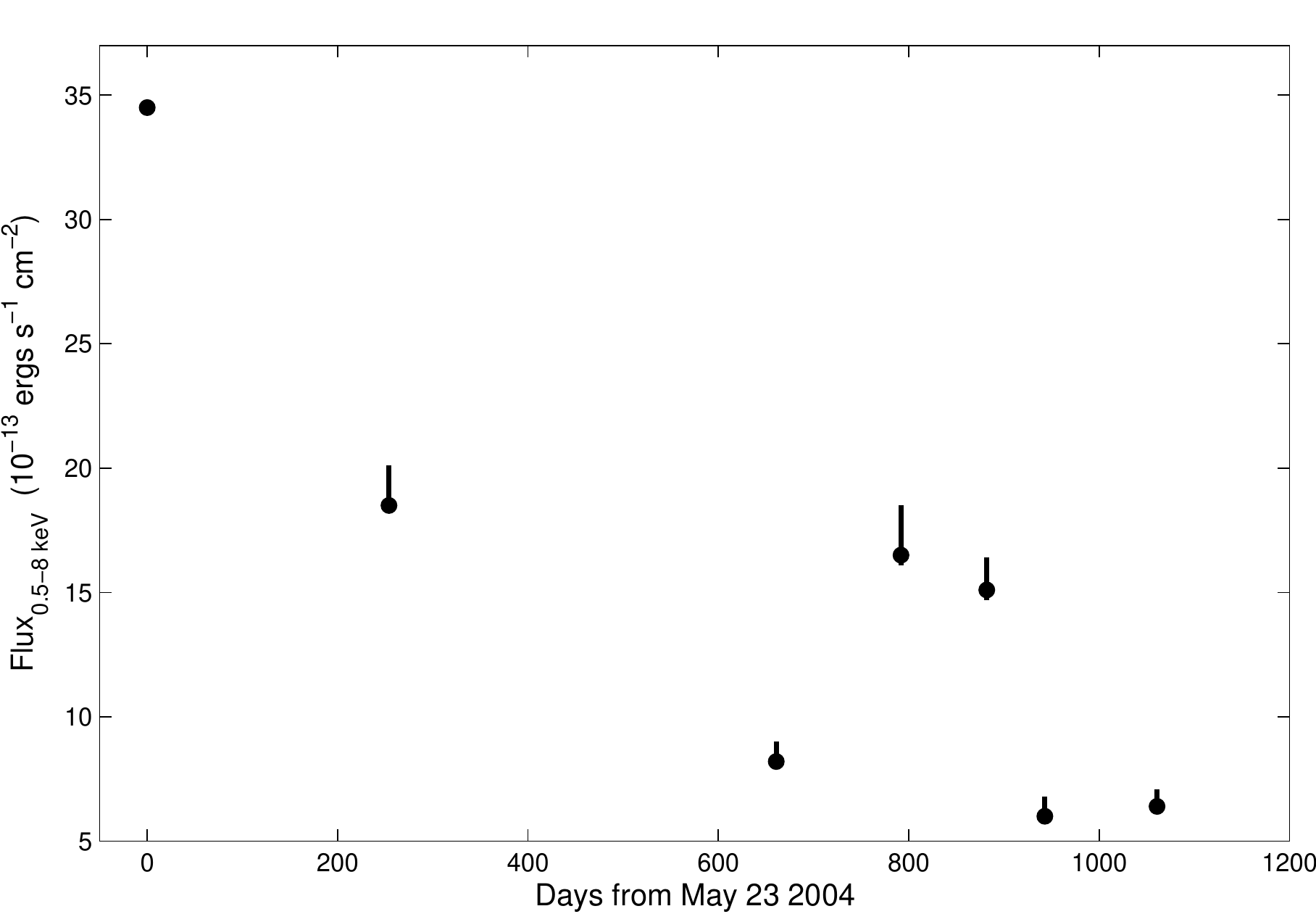}}
\caption{The long term X--ray variability of \src.}
\label{longtermflux}
\end{figure}

At  radio wavelengths,  \src\  does not  show significant  inter--year
variability   at   2   or   3.6~cm   \citep{nagar02aap}.    At   6~cm,
\citet{giroletti05apj:ngc4278} discovered modest variability on yearly
time--scales with the flux decreasing by a factor of 10\% between 1995
and  2000.  At  UV wavelengths,  \citet{ngc4278:cardullo09} discovered
that  \src\ is  variable on  months timescales  showing that  its flux
increased $\sim$2 times between two different \hst\ observations. This
is  in  agreement  with  \citet{maoz05apj:linervarUV}  who  discovered
variability in 12 of 16  LINERsI/LLAGN objects observed with the \hst\
ACS/HRC F250W and F330W filters.  Nine of the 12 galaxies that show UV
variability  in their sample  have detected  radio cores  suggesting a
connection  between UV variability  and the  presence of  radio cores.
\src\ enforces this  correspondence as it shows a  luminous radio core
as  well  as  significant  UV variability.   Additionally,  we  detect
optical  nuclear variability  on years  time--scale as  the  fluxes we
derive from \hst\ ACS F475W and F850LP filters are 2 to 4 times higher
than optical fluxes  derived from \hst\ WFPC2 F555W  and F814W filters
(we do not discard optical  variability on shorter time--scales but no
intra--year optical observations were ever made for \src). In X--rays,
\src\  show   no  significant   short  time--scale  (hours   to  days)
variability  during  the  different  \chandra\ observations  with  the
``excess variance'', $\sigma^{2}_{rms}$, never exceeding 0.01, with an
error $<15$\%, which  means that \src\ does not  follow the same trend
as  Seyfert  galaxies  showing  stronger variability  with  decreasing
luminosity  \citep{nandra97apj:variance}.  On  the other  hand, during
the  \xmm\ observation, \src\  exhibits short--time  scale variability
with the  flux increasing  by a factor  of 10\%\  on a 1  hour period.
Moreover,  a significant monthly  time--scale variability  is observed
between the different \chandra\ observations and between \chandra\ and
\xmm\ observations (Fig.~\ref{longtermflux})  where the flux increased
by a factor of $\sim$3 on a  few months time--scale and by a factor of
5  between the  faintest and  the brightest  observation  separated by
$\sim$3 years.  \citet{ptak98apj:variance}  suggested that the lack of
variability on  short ($<$  1 day) time--scales  and the  detection of
variability on longer  time--scales could be due to  the fact that the
regions responsible  for the X-ray  emission are larger in  LINERs and
LLAGN than in  typical Seyfert galaxies.  One possibility  is that the
X-ray emission  originates from an optically  thin geometrically thick
flow (RIAF) rather than from a geometrically thin accretion disk.

This  scenario  could  be   present  during  the  different  \chandra\
observations  corresponding to  the lowest  flux level.   The  lack of
variability  on  time--scales  of  hours  to  days  found  during  the
\chandra\  observation  could  be  compatible with  a  large  emission
region,  such  as an  extended  RIAF  emission.   However, in  AGN  an
anti--correlation exists between the  2--10 keV X--ray variability and
the mass  of the  black hole \citep{lu01mnras:varmassbhrel}  with some
exceptions     whenever     the     accretion     rate     is     high
\citep[e.g.,][]{reeves02mnras:pds456}.  Hence, the lack of variability
during the \chandra\  observations in NGC 4278 is  consistent with the
high     black     hole     mass    of     $3\times10^{8}$~M$_{\odot}$
\citep{wang03MNRAS:Mbh,chiaberge05ApJ:Mbh} and the low accretion rate.

The short time--scale  variability (t$\sim$1.5~h), observed during the
\xmm\ observation (where  the X-ray flux level was  highest), could be
explained by a  disturbance in the (extended) RIAF  or by the presence
of a more compact region, e.g., an accretion disk truncated at a lower
radius.  In the  context of  X--ray fluctuations  in a  RIAF scenario,
\citet{manmoto96apj:adafvar}  and  \citet{takeuchi97apj:adafvar}  were
able  to explain  the X--ray  shots  seen in  Cygnus~X-1 light  curves
\citep{negoro94apj:cygx1} as due to a disturbance in an extended RIAF.
However, contrarily to Cygnus~X-1 that switches from a soft state to a
hard state just  after the peak, \src\ remains in  a soft state during
the whole observation (including  the variation). Additionally, if the
variation in  the case  of the  \xmm\ observation is  the result  of a
disturbance  in  the  accretion   flow,  the  maximum  radius  can  be
determined from the time duration of the variation from the free--fall
time--scale \citep{manmoto96apj:adafvar}:

\begin{equation}
T(ff)\approx4\times\left(\frac{r}{10^{3}r_{g}}\right)^{3/2}\left(\frac{M}{10M_{\odot}}\right)s
\end{equation}

where $M$ is the BH mass,  $T(ff)$ is the free-fall time (the duration
of  the  shot), $r$  is  the transition  radius,  and  $r_{g}$ is  the
gravitational radius, $GM/c^{2}$.  This  yields to a transition radius
of  $r\sim1.3~r_{g}$ between  the RIAF  and the  classical  thin disk.
Therefore a disturbance in an extended RIAF region appears unlikely to
explain the  short time--scale variation observed  from \src, although
further works on  variability from RIAF are required  to establish any
definitive  conclusion.  Therefore,  the  presence of  a more  compact
region  during the  \xmm\ observation,  (i.e.  the  accretion  disk is
similar to that observed in  low luminosity Seyfert~1 galaxies or most
probably truncated  at a lower  radius than that during  the \chandra\
observations) appears to  be a more viable explanation  for this short
time--scale variability.

%that the region responsible for the X--ray emission is more compact.

%Therefore during  the high state flux,  \src\ nucleus seems  not to be
%compatible  with  an  {\it  extended} \textbf{RIAF}  structure  and  a
%truncated disk  at a large  radius.  Consequently, the  only plausible
%explanation could  be that during the \xmm\  observation the accretion
%disk was similar to that  observed in low luminosity Seyfert~1 or most
%probably truncated  at a lower  radius than that during  the \chandra\
%observations.}

\subsection{\src\ SED comparison to other LLAGN}

We now turn  to examine whether the X-ray  continuum emission of \src\
originates  from  a   classical  optically  thick  geometrically  thin
accretion disk  or from a RIAF  structure forming around  a black hole
(BH) whenever  the radiative efficiency  is very low.   The bolometric
luminosity L$_{bol}$ is estimated by integrating the SED from radio to
X-ray  data, neglecting upper  limits such  as points  in the  IR.  As
mentioned in  sect.~5.2, \src\ as  most of LINERsI/LLAGN, show  a high
degree  of variability  on months  timescales.  Hence,  two bolometric
luminosities were  derived where the highest X--ray  flux (\xmm\ flux)
is used  to derive  the highest $L_{bol}$  and the lowest  X--ray flux
(\chandra\  obs.   ID 7080  flux)  is  used  to calculate  the  lowest
$L_{bol}$.          This        gives         L$_{bol}$        between
$2.21\times10^{41}$~erg~s$^{-1}$         (low        state)        and
$2.88\times10^{41}$~erg~s$^{-1}$  (high  state)   which  leads  to  an
Eddington  ratio   $L_{bol}/L_{Edd}$  between  5.7$\times10^{-6}$  and
7.4$\times10^{-6}$.  Similar $L_{bol}/L_{Edd}$  value were derived for
the LINER galaxy M81 \citep{pianmnras10}.  Two possibilities emerge to
explain  the  faint luminosities  in  both  the  low and  high  state;
accretion is  occurring at sub-Eddington rate and/or  the accretion is
radiatively inefficient.

\citet{maoz07MNRAS}, when studying a sample of 13 nearby galaxies with
LINER nuclei,  argued that  their SED do  resemble the SEDs  of normal
Seyfert  galaxies, and  hence the  mode of  accretion, ``geometrically
thin optically thick accretion  flow'', and the X-ray continuum, ``the
result  of  inverse compton  scattering  of  UV/soft  X-ray in  a  hot
optically thin  plasma in  a corona'', are  also similar.   The author
based  his  arguments  on   two  parameters:  the  optical  to  X-ray,
$\alpha_{ox}$, slope  (calculated using  the 2500~\AA\ flux),  and the
radio  loudness, $R_{UV}$,  (also based  on the  2500~\AA\  flux).  He
showed that  the $\alpha_{ox}$ values  derived for his  sample overlap
with   the  values   derived   for  a   sample   of  broad-lined   AGN
\citep{steffen06AJ:aox}    and   demonstrated    that    even   though
LINERsI/LLAGN are louder by a factor of $\sim$100 than high-luminosity
quasars, most  of the  LINERs are considered  radio quiet  compared to
their  2500~\AA\ flux.   \src\  $\alpha_{ox}$ is  -1.2 (see  sect.~4),
comparable  to the  values derived  for \citet{maoz07MNRAS}  sample of
LLAGN and it follows the extension to low luminosities of the relation
between the  $\alpha_{ox}$ and the 2500~\AA\ luminosity  derived for a
sample of  Seyfert galaxies \citep{steffen06AJ:aox}.   Moreover, \src\
$R_{UV}$  is 2.7,  thus it  stands in  the radio  quiet branch  of the
relation between $R_{UV}$ and the 2500~\AA\ luminosity for a sample of
luminous AGN \citep{sikora07apj:radloudagn,maoz07MNRAS}.  We note that
our  results may be  slightly biased  by the  fact that  the 2500~\AA\
luminosity  we derive  for \src\  comes  from the  UVM2 \xmm\  optical
monitor telescope  where neither the peak $\lambda$  nor the effective
$\lambda$ are as close as  the \hst\ ACS/HRC F250W filter to 2500~\AA\
used by \citet{maoz07MNRAS}.

A RIAF model  has been suggested as a solution  for the low luminosity
observed   in   a   number   of  LINERsI/LLAGN;   M81   and   NGC~4579
\citep{quataert99apj:m81ngc4579},                              NGC~4258
\citep{gammie99apj:ngc4258},    NGC~3998    \citep{ptak04apj:ngc3998},
NGC~1097 \citep{nemmen06apj:ngc1097}, and, more recently, for a sample
of  24  LINER sources  \citep{nemmen10:lineradaf}.   Authors of  these
papers emphasize  the fact that SEDs  of LLAGN in general  are in fact
different than normal Seyfert SEDs.   They differ in a number of ways,
mainly, the lack of the canonical ``big blue bump'' at UV wavelengths,
observed in almost  all of the luminous galaxies,  which is associated
with  thermal  emission from  a  geometrically  thin, optically  thick
accretion disk  \citep{ho05apss,ho02aspcs}. Instead, a  maximum in the
SED is peaking in the mid-IR.   The presence of radio jets is a common
feature  in  LLAGN  as  for  they  tend  to  be  radio  loud  systems.
Additionally,  the lack  of a  broad Fe~K${\alpha}$  emission  line at
6.4~keV in most of the LLAGN  studied up until now, originating from a
standard disk in typical luminous AGN, might suggest the truncation of
this  structure at  a radius  much  larger than  the innermost  stable
circular orbit  (ISCO) where a radiatively  inefficient accretion flow
(RIAF)  structure forms  in the  inner part.   We note  that  the last
argument, lack of a broad Fe K line, used to point at the existence of
a RIAF in  LINERs is not decisive since a  small fraction, 25-30\%, of
Seyfert  galaxies  exhibit  broadened  emission  in the  iron  K  band
\citep{nandra07mnras:felinesey}.

\begin{figure}[!t]
\centerline{\includegraphics[angle=0,width=0.5\textwidth]{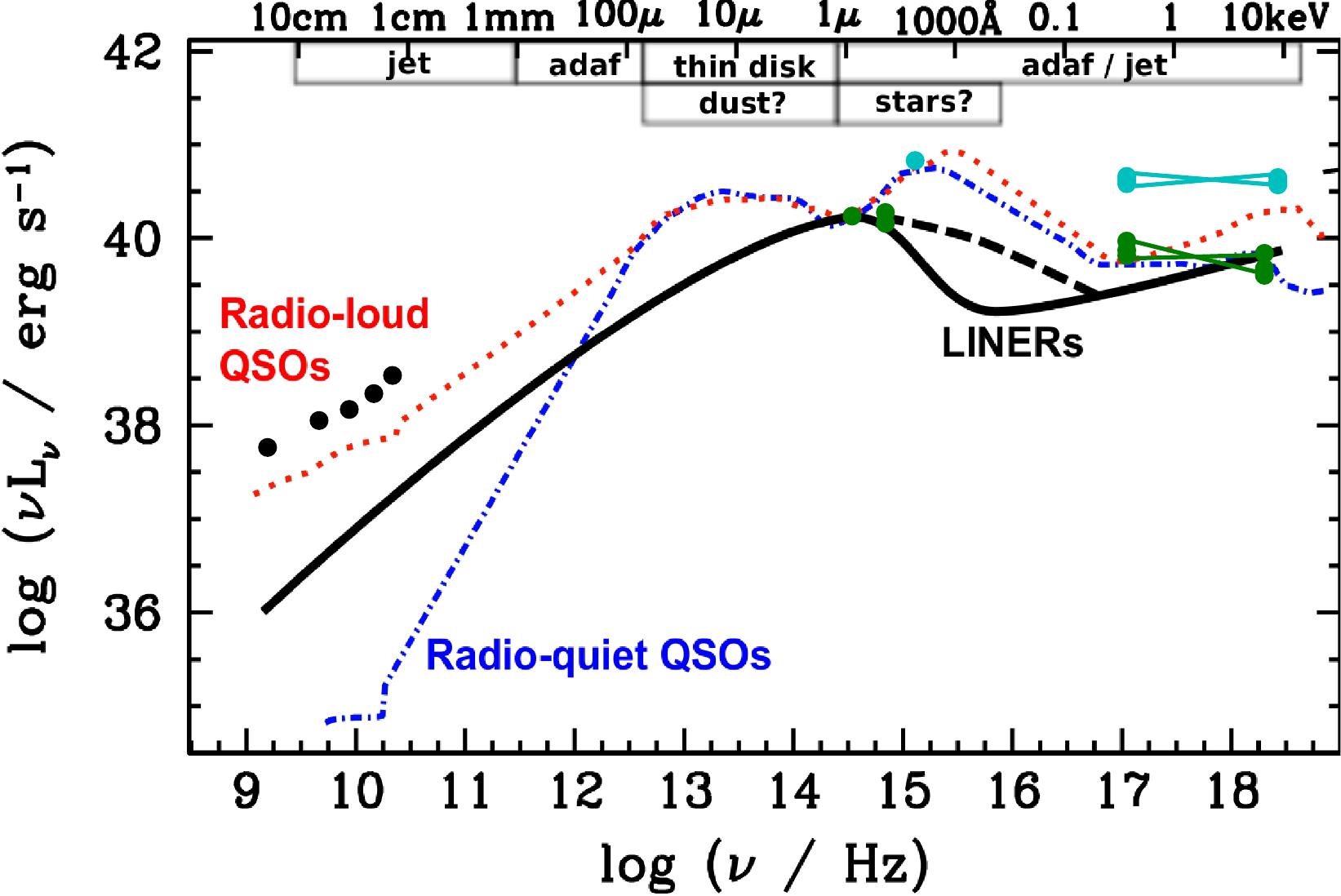}}
\caption{\src\ SED: black dots represent the radio data. Green dots represent the contemporary \hst/ACS and the \chandra\ 7081 observation fluxes. The light--blue dots correspond to the simultaneous \xmm\ OM and pn fluxes. The average SED of a sample of LINERs is shown as solid black line \citep[figure adopted from][]{nemmen10:lineradaf}. The dashed line represents the effect of extinction correction. For comparison, the average SED of \citet{elvis94apjs:quasar} for radio-loud (dotted line) and radio-quiet (dot--dashed line) quasars. All SEDs are normalized to 1$\mu$m.}
\label{fullsed}
\end{figure}

The absence of a narrow Fe~K${\alpha}$ emission line at 6.4~keV with a
22~eV upper limit on its EW  is in agreement with the possibility of a
RIAF presence at the center of \src. However, the large upper limit of
118~eV on the EW of a broad line at 6.4~keV is not discriminative.  In
order to examine  more accurately the possibility of  a RIAF structure
and to identify  the accretion mode occurring at  the nucleus of \src\
we       plotted       in      Fig.~\ref{fullsed}       \citep[adopted
from][]{nemmen10:lineradaf} fluxes derived for \src\ from simultaneous
or  close--by  observations.   The  black solid  line  represents  the
average  SED  of a  sample  of LINER  nuclei  (the  dashed black  line
represents  the effect  of extinction),  the dotted  red line  and the
dot--dashed    blue    line   represents    the    average   SED    of
\citet{elvis94apjs:quasar}  sample  of  radio--loud  and  radio--quiet
quasars,   respectively.   All   SEDs  are   normalized   to  1$\mu$m.
Light--blue  dots represent  the simultaneous  \xmm\ UVM2  flux  at UV
wavelengths  and  the  EPIC-pn  fluxes  at X--rays.   The  green  dots
represent the  ACS/WFC optical  data and the  \chandra\ obs.   ID 7081
X--ray  fluxes.   The  EPIC-pn  and  the OM  observations  were  taken
simultaneously and  the time difference between the  \chandra\ and the
\hst\ data is less than  two months and therefore assumed simultaneous
as  large   optical  and  X--ray  variability  is   observed  only  on
time--scales of several months  (see sect.~5.2).  Black dots represent
the radio data  and are considered constant on  time--scale of several
years.

If  we compare  the \src\  SED during  the low  \chandra\  X--ray flux
(black  and  green   dots  combined)  to  the  other   SEDs  shown  in
Fig.~\ref{fullsed},  we  can consider  \src\  as  a  radio loud  LINER
galaxy.  The radio emission is more important than the one observed in
radio loud  quasars, plus, optical and  X--ray luminosities correspond
to the  luminosities derived for the  sample of LINERs.   On the other
hand, the  \src\ SED  during the \xmm\  observation, where  the X--ray
flux is the highest, (black and light--blue dots combined) seems to be
similar  to the SEDs  of a  small number  of low  luminosity Seyfert~1
galaxies,  in   which  the   UV  to  X--ray   slope  is   almost  flat
\citep{papadakis08aa:sey1SED},  where  the  authors  showed  that  the
accretion    mode    in     low    luminosity    Seyfert~1    galaxies
($L_{X}/L_{Edd}<10^{-4}$)  resemble  the  accretion mode  detected  in
luminous AGN.

As  \src\ shows  a high  degree of  variability on  months time--scale
fluctuating apparently between a low--state SED (black and green dots)
and  a  high--state  SED  (black  and light--blue  dots)  where  short
time--scale  variability is  detected, we  consider that  the emission
mechanism responsible  for the  bulk of energy  from radio  to X--rays
during the low state is jet/RIAF dominated; a thin accretion disk most
probably  truncated  but  at  a  lower radius  than  that  during  the
low--state  level  would be  responsible  for  the  optical to  X--ray
emission during the high--state SED.

%Therefore, \src\ could be a intermediate LINER--Seyfert galaxy.

%\footnote{We note that regardless of  whether \src\ is in a high state
%or in a low state, the  source remains in a soft state.  Therefore, we
%are  not  assuming the  disappearance  of  the  jet/ADAF structure  or
%emission  when  \src\  SED  is  comparable to  Seyfert  SEDs,  we  are
%considering the  ADAF structure  to shrink and  the thin  accretion to
%come closer to the nucleus to a point where the emission from the thin
%accretion disk would be more relevant.}

\section{Summary and conclusion}

We  have studied  in detail  seven  X-ray observations  for \src,  six
ACIS-S  \chandra\  observations   plus  one  \xmm\  observation.   The
observations  covered  a  three  year period  allowing  an  extensive
variability study. No short time--scale (minutes to hours) variability
is detected during the \chandra\ X--ray observations, while during the
\xmm\ observation, where the highest  flux level is observed, the flux
increases by  a factor  of 10\%\  in a 1  hour period.   A significant
months  time--scale  variability  is  observed between  the  different
\chandra\  observations and between  \chandra\ and  \xmm\ observations
where  the flux  increased by  a  factor of  $\sim$3 on  a few  months
time--scale  and  by a  factor  of 5  between  the  faintest and  the
brightest  observations  separated by  $\sim$3  years.  We checked  the
validity of increasing variability with decreasing luminosity detected
in Seyfert  galaxies \citep{nandra97apj:variance}, \src,  like most of
LINER     sources,    does     not    follow     the     same    trend
\citep{ptak98apj:variance}.   However, based on  the anti--correlation
relation between variability and the  mass of the black hole with some
exceptions     whenever     the     accretion     rate     is     high
\citep{reeves02mnras:pds456},   no  short  time--scale   variation  is
expected in \src\ as  it harbors a $\sim3\times10^8$~M$_{\odot}$ black
hole  and  accretes  at  low  rate.  However,  the  short--time  scale
variability  detected during  the  \xmm\ observation  might suggest  a
change in the  central engine of \src\ at  that point.  No variability
is detected in  the hardness ratio during the  observations being soft
in all cases.

The best spectral  fit found for the six  \chandra\ observations is an
absorbed power  law plus a thermal component.   The intrinsic hydrogen
column  density  affecting  the  power  law  has  an  upper  limit  of
$6.7\times10^{20}$~cm$^{-2}$, while the  average photon index $\Gamma$
is $2.2_{-0.2}^{+0.1}$.   The temperature of the  thermal component is
$0.63_{-0.04}^{+0.05}$~keV.   Because of  the  \xmm\ moderate  angular
resolution  compared  to  \chandra,  we subtracted,  in  a  10\arcsec\
radius,  the contribution  from  X-ray binaries  and diffuse  emission
calculated  from  the  \chandra\   observations.   We  find  that  the
power--law  emission is dominant  over the  whole 0.5--8~keV  band and
therefore no thermal component is needed.  The hydrogen column density
is $(3.8\pm0.8)\times10^{20}$~cm$^{-2}$  affecting a power  law with a
photon  index  $\Gamma$  of  2.1.  No
Fe~K${\alpha}$ emission line at 6.4~keV is detected with a 22~eV upper
limit on its equivalent width.

We        measured       a        UV        flux       of        about
$1.2\times10^{-15}$~ergs~cm$^{-2}$~s$^{-1}$~\AA$^{-1}$  from the \xmm\
optical  monitor  observation  of  \src.   The  advantage  of  such  a
measurement is  the simultaneity  with the X--ray  observation crucial
for  comparing  such  sources  with  normal galaxies  as  \src\  shows
significant variability on months time--scales.  We calculated optical
fluxes  coming from \hst\  ACS observations  in two  different filters
(F475W and F850LP).  These fluxes are 2 to 4 times higher than optical
fluxes, coming from the \hst\  WFPC2 F555W and F814W filters, observed
$\sim12$ years  earlier.  Our optical  fluxes are contemporary  to the
\chandra\  fluxes  coming  from   the  7081  observation  as  the  two
observations are separated by less than two months.

In order to determine the origin of the emission mechanism responsible
for  the bulk  of  energy from  radio  to X--rays  and  to assess  the
geometry of  the central engine in  \src\, we plotted  the SED derived
from simultaneous  observations (radio\footnote{We consider  the radio
flux constant,  and include it in the  simultaneous observations since
it  does   not  show  significant  variability  on   a  several  years
time--scale         (see         sect.~5.2).}$+$UV$+$X-rays        and
radio$+$optical$+$X--rays) with  the SED of a sample  of LINER sources
\citep{eracleous10:linersed}  along with radio--quiet  and radio--loud
quasars \citep[fig.~8;  adopted from][]{nemmen10:lineradaf}.  As \src\
appears to fluctuate in  optical, UV and X--rays on month--timescales,
it  appears that  \src\ SED  is consistent  with LINER  SEDs  when the
X--ray emission is weak and therefore we suppose a jet/RIAF origin for
the radio to X--ray bulk of energy. On the other hand, when the X--ray
emission is  important, the  \src\ SED  is similar to  the SED  of low
luminosity Seyfert  galaxies and a thin accretion  disk is responsible
for the bulk  of energy from UV to X--rays.   Hence, we consider \src\
to  alternate  between LINER-like  and  Seyfert-like nuclear  activity
depending on the strength of its X--ray emission.

\acknowledgements This research has made use of the data obtained from
the  \textsl{Chandra  Data  Archive}  and the  \textsl{Chandra  Source
Catalog}, and  software provided by the  \textsl{Chandra X-ray Center}
(CXC) in the application packages  CIAO and Chips.  This work is based
on observations with \xmm, an ESA science mission with instruments and
contributions directly funded by ESA Member States and the USA (NASA).
This research  has made use of  the SIMBAD database,  operated at CDS,
Strasbourg,  France.  This  research  has made  use  of the  NASA/IPAC
Extragalactic Database  (NED) which is operated by  the Jet Propulsion
Laboratory,  California Institute of  Technology, under  contract with
the National Aeronautics and Space Administration.  G.Y. would like to
thank Patrick Broos for helpful comments on the use of the AE package.
The authors would also like to thank the referee for fruitful comments
that improved the quality of the manuscript.

%------------------------------------
%       References
%------------------------------------

%\bibliographystyle{aa} \bibliography{mybib}

\begin{thebibliography}{77}
\expandafter\ifx\csname natexlab\endcsname\relax\def\natexlab#1{#1}\fi

\bibitem[{{Alonso-Herrero} {et~al.}(2000){Alonso-Herrero}, {Rieke}, {Rieke}, \&
  {Shields}}]{alonso-herrero00ApJ:starburstinliners}
{Alonso-Herrero}, A., {Rieke}, M.~J., {Rieke}, G.~H., \& {Shields}, J.~C. 2000,
  \apj, 530, 688

\bibitem[{{Arnaud}(1996)}]{arnaud96conf}
{Arnaud}, K.~A. 1996, in ASP Conf. Ser. 101: Astronomical Data Analysis
  Software and Systems V, 17

\bibitem[{{Brassington} {et~al.}(2009){Brassington}, {Fabbiano}, {Kim},
  {Zezas}, {Zepf}, {Kundu}, {Angelini}, {Davies}, {Gallagher}, {Kalogera},
  {Fragos}, {King}, {Pellegrini}, \& {Trinchieri}}]{brassington09apjs}
{Brassington}, N.~J., {Fabbiano}, G., {Kim}, D.-W., {et~al.} 2009, \apjs, 181,
  605

\bibitem[{{Brenneman} {et~al.}(2009){Brenneman}, {Weaver}, {Kadler}, {Tueller},
  {Marscher}, {Ros}, {Zensus}, {Kovalev}, {Aller}, {Aller}, {Irwin}, {Kerp}, \&
  {Kaufmann}}]{brenneman09ApJ:ngc1052}
{Brenneman}, L.~W., {Weaver}, K.~A., {Kadler}, M., {et~al.} 2009, \apj, 698,
  528

\bibitem[{{Broos} {et~al.}(2010){Broos}, {Townsley}, {Feigelson}, {Getman},
  {Bauer}, \& {Garmire}}]{broos2010AE}
{Broos}, P.~S., {Townsley}, L.~K., {Feigelson}, E.~D., {et~al.} 2010, ArXiv
  e-prints

\bibitem[{{Broos} {et~al.}(1998){Broos}, {Townsley}, \& {Nousek}}]{broos98pile}
{Broos}, P.~S., {Townsley}, L.~K., \& {Nousek}, J.~A. 1998, in Society of
  Photo-Optical Instrumentation Engineers (SPIE) Conference Series, Vol. 3444,
  Society of Photo-Optical Instrumentation Engineers (SPIE) Conference Series,
  ed. {R.~B.~Hoover \& A.~B.~Walker}, 30--35

\bibitem[{{Capetti} {et~al.}(2002){Capetti}, {Celotti}, {Chiaberge}, {de
  Ruiter}, {Fanti}, {Morganti}, \& {Parma}}]{capetti02aap}
{Capetti}, A., {Celotti}, A., {Chiaberge}, M., {et~al.} 2002, \aap, 383, 104

\bibitem[{{Capetti} {et~al.}(2000){Capetti}, {de Ruiter}, {Fanti}, {Morganti},
  {Parma}, \& {Ulrich}}]{capetti00aa:hst}
{Capetti}, A., {de Ruiter}, H.~R., {Fanti}, R., {et~al.} 2000, \aap, 362, 871

\bibitem[{{Cardelli} {et~al.}(1989){Cardelli}, {Clayton}, \&
  {Mathis}}]{cardelli89apj:gal.ext}
{Cardelli}, J.~A., {Clayton}, G.~C., \& {Mathis}, J.~S. 1989, \apj, 345, 245

\bibitem[{{Cardullo} {et~al.}(2009){Cardullo}, {Corsini}, {Beifiori}, {Buson},
  {Dalla Bont{\~a}}, {Morelli}, {Pizzella}, \& {Bertola}}]{ngc4278:cardullo09}
{Cardullo}, A., {Corsini}, E.~M., {Beifiori}, A., {et~al.} 2009, \aap, 508, 641

\bibitem[{{Chiaberge} {et~al.}(2005){Chiaberge}, {Capetti}, \&
  {Macchetto}}]{chiaberge05ApJ:Mbh}
{Chiaberge}, M., {Capetti}, A., \& {Macchetto}, F.~D. 2005, \apj, 625, 716

\bibitem[{{Davis}(2001)}]{davis01apjpile}
{Davis}, J.~E. 2001, \apj, 562, 575

\bibitem[{{de Vaucouleurs} {et~al.}(1991){de Vaucouleurs}, {de Vaucouleurs},
  {Corwin}, {Buta}, {Paturel}, \& {Fouque}}]{devaucouleurs91}
{de Vaucouleurs}, G., {de Vaucouleurs}, A., {Corwin}, Jr., H.~G., {et~al.}
  1991, {Third Reference Catalogue of Bright Galaxies}, ed. {de Vaucouleurs,
  G., de Vaucouleurs, A., Corwin, H.~G., Jr., Buta, R.~J., Paturel, G., \&
  Fouque, P.}

\bibitem[{{Elvis} {et~al.}(1994){Elvis}, {Wilkes}, {McDowell}, {Green},
  {Bechtold}, {Willner}, {Oey}, {Polomski}, \& {Cutri}}]{elvis94apjs:quasar}
{Elvis}, M., {Wilkes}, B.~J., {McDowell}, J.~C., {et~al.} 1994, \apjs, 95, 1

\bibitem[{{Eracleous} {et~al.}(2010){Eracleous}, {Hwang}, \&
  {Flohic}}]{eracleous10:linersed}
{Eracleous}, M., {Hwang}, J.~A., \& {Flohic}, H.~M.~L.~G. 2010, \apjs, 187, 135

\bibitem[{{Fabbiano}(2006)}]{fabbiano06aa:lmxbpop}
{Fabbiano}, G. 2006, \araa, 44, 323

\bibitem[{{Flohic} {et~al.}(2006){Flohic}, {Eracleous}, {Chartas}, {Shields},
  \& {Moran}}]{flohic06apj}
{Flohic}, H.~M.~L.~G., {Eracleous}, M., {Chartas}, G., {Shields}, J.~C., \&
  {Moran}, E.~C. 2006, \apj, 647, 140

\bibitem[{{Freeman} {et~al.}(2002){Freeman}, {Kashyap}, {Rosner}, \&
  {Lamb}}]{wavdetect:freeman02apjs}
{Freeman}, P.~E., {Kashyap}, V., {Rosner}, R., \& {Lamb}, D.~Q. 2002, \apjs,
  138, 185

\bibitem[{{Gammie} {et~al.}(1999){Gammie}, {Narayan}, \&
  {Blandford}}]{gammie99apj:ngc4258}
{Gammie}, C.~F., {Narayan}, R., \& {Blandford}, R. 1999, \apj, 516, 177

\bibitem[{{Getman} {et~al.}(2005){Getman}, {Flaccomio}, {Broos}, {Grosso},
  {Tsujimoto}, {Townsley}, {Garmire}, {Kastner}, {Li}, {Harnden}, {Wolk},
  {Murray}, {Lada}, {Muench}, {McCaughrean}, {Meeus}, {Damiani}, {Micela},
  {Sciortino}, {Bally}, {Hillenbrand}, {Herbst}, {Preibisch}, \&
  {Feigelson}}]{getman05apjsAE}
{Getman}, K.~V., {Flaccomio}, E., {Broos}, P.~S., {et~al.} 2005, \apjs, 160,
  319

\bibitem[{{Giroletti} {et~al.}(2005){Giroletti}, {Taylor}, \&
  {Giovannini}}]{giroletti05apj:ngc4278}
{Giroletti}, M., {Taylor}, G.~B., \& {Giovannini}, G. 2005, \apj, 622, 178

\bibitem[{{Gonz{\'a}lez-Mart{\'{\i}}n}
  {et~al.}(2009){Gonz{\'a}lez-Mart{\'{\i}}n}, {Masegosa}, {M{\'a}rquez},
  {Guainazzi}, \& {Jim{\'e}nez-Bail{\'o}n}}]{gonzalezmartin09aa}
{Gonz{\'a}lez-Mart{\'{\i}}n}, O., {Masegosa}, J., {M{\'a}rquez}, I.,
  {Guainazzi}, M., \& {Jim{\'e}nez-Bail{\'o}n}, E. 2009, \aap, 506, 1107

\bibitem[{{Guainazzi} {et~al.}(2005){Guainazzi}, {Matt}, \&
  {Perola}}]{guainazzi05aa:sey2}
{Guainazzi}, M., {Matt}, G., \& {Perola}, G.~C. 2005, \aap, 444, 119

\bibitem[{{Heckman}(1980)}]{heckman80aap}
{Heckman}, T.~M. 1980, \aap, 87, 152

\bibitem[{{Heckman} {et~al.}(1983){Heckman}, {Lebofsky}, {Rieke}, \& {van
  Breugel}}]{heckman83apj:irobs}
{Heckman}, T.~M., {Lebofsky}, M.~J., {Rieke}, G.~H., \& {van Breugel}, W. 1983,
  \apj, 272, 400

\bibitem[{{Ho}(1999)}]{ho99sed}
{Ho}, L.~C. 1999, \apj, 516, 672

\bibitem[{{Ho}(2002)}]{ho02aspcs}
{Ho}, L.~C. 2002, in Astronomical Society of the Pacific Conference Series,
  Vol. 258, Issues in Unification of Active Galactic Nuclei, ed. {R.~Maiolino,
  A.~Marconi, \& N.~Nagar}, 165--+

\bibitem[{{Ho}(2005)}]{ho05apss}
{Ho}, L.~C. 2005, \apss, 300, 219

\bibitem[{{Ho}(2008)}]{ho08aa:review}
{Ho}, L.~C. 2008, \araa, 46, 475

\bibitem[{{Ho} {et~al.}(2001){Ho}, {Feigelson}, {Townsley}, {Sambruna},
  {Garmire}, {Brandt}, {Filippenko}, {Griffiths}, {Ptak}, \&
  {Sargent}}]{ho01apjl}
{Ho}, L.~C., {Feigelson}, E.~D., {Townsley}, L.~K., {et~al.} 2001, \apjl, 549,
  L51

\bibitem[{{Ho} {et~al.}(1993){Ho}, {Filippenko}, \&
  {Sargent}}]{ho93ApJ:linerAGN}
{Ho}, L.~C., {Filippenko}, A.~V., \& {Sargent}, W.~L.~W. 1993, \apj, 417, 63

\bibitem[{{Ho} {et~al.}(1997){Ho}, {Filippenko}, {Sargent}, \&
  {Peng}}]{ho97apjs}
{Ho}, L.~C., {Filippenko}, A.~V., {Sargent}, W.~L.~W., \& {Peng}, C.~Y. 1997,
  \apjs, 112, 391

\bibitem[{{Jim{\'e}nez-Bail{\'o}n} {et~al.}(2005){Jim{\'e}nez-Bail{\'o}n},
  {Piconcelli}, {Guainazzi}, {Schartel}, {Rodr{\'{\i}}guez-Pascual}, \&
  {Santos-Lle{\'o}}}]{jiminez05AA:pgquasarfeline}
{Jim{\'e}nez-Bail{\'o}n}, E., {Piconcelli}, E., {Guainazzi}, M., {et~al.} 2005,
  \aap, 435, 449

\bibitem[{{Jones} {et~al.}(1984){Jones}, {Wrobel}, \&
  {Shaffer}}]{jones84apj:4278radobs}
{Jones}, D.~L., {Wrobel}, J.~M., \& {Shaffer}, D.~B. 1984, \apj, 276, 480

\bibitem[{{Kalberla} {et~al.}(2005){Kalberla}, {Burton}, {Hartmann}, {Arnal},
  {Bajaja}, {Morras}, \& {P{\"o}ppel}}]{kalberla05aa:nh}
{Kalberla}, P.~M.~W., {Burton}, W.~B., {Hartmann}, D., {et~al.} 2005, \aap,
  440, 775

\bibitem[{{Kim} \& {Fabbiano}(2004)}]{kim04ApJlmxb}
{Kim}, D. \& {Fabbiano}, G. 2004, \apj, 611, 846

\bibitem[{{Koratkar} \& {Blaes}(1999)}]{koratkar99pasp:seybbb}
{Koratkar}, A. \& {Blaes}, O. 1999, \pasp, 111, 1

\bibitem[{{Lauer} {et~al.}(2005){Lauer}, {Faber}, {Gebhardt}, {Richstone},
  {Tremaine}, {Ajhar}, {Aller}, {Bender}, {Dressler}, {Filippenko}, {Green},
  {Grillmair}, {Ho}, {Kormendy}, {Magorrian}, {Pinkney}, \&
  {Siopis}}]{lauer05aj:hst}
{Lauer}, T.~R., {Faber}, S.~M., {Gebhardt}, K., {et~al.} 2005, \aj, 129, 2138

\bibitem[{{Longmore} \& {Sharples}(1982)}]{longmore82mnras:irobs}
{Longmore}, A.~J. \& {Sharples}, R.~M. 1982, \mnras, 201, 111

\bibitem[{{Lu} \& {Yu}(2001)}]{lu01mnras:varmassbhrel}
{Lu}, Y. \& {Yu}, Q. 2001, \mnras, 324, 653

\bibitem[{{Malkan} \& {Sargent}(1982)}]{malkan82apj:uvexcSey1}
{Malkan}, M.~A. \& {Sargent}, W.~L.~W. 1982, \apj, 254, 22

\bibitem[{{Manmoto} {et~al.}(1996){Manmoto}, {Takeuchi}, {Mineshige},
  {Matsumoto}, \& {Negoro}}]{manmoto96apj:adafvar}
{Manmoto}, T., {Takeuchi}, M., {Mineshige}, S., {Matsumoto}, R., \& {Negoro},
  H. 1996, \apjl, 464, L135+

\bibitem[{{Maoz}(2007)}]{maoz07MNRAS}
{Maoz}, D. 2007, \mnras, 377, 1696

\bibitem[{{Maoz} {et~al.}(2005){Maoz}, {Nagar}, {Falcke}, \&
  {Wilson}}]{maoz05apj:linervarUV}
{Maoz}, D., {Nagar}, N.~M., {Falcke}, H., \& {Wilson}, A.~S. 2005, \apj, 625,
  699

\bibitem[{{Mason} {et~al.}(2001){Mason}, {Breeveld}, {Much}, {Carter},
  {Cordova}, {Cropper}, {Fordham}, {Huckle}, {Ho}, {Kawakami}, {Kennea},
  {Kennedy}, {Mittaz}, {Pandel}, {Priedhorsky}, {Sasseen}, {Shirey}, {Smith},
  \& {Vreux}}]{mason01aa:om}
{Mason}, K.~O., {Breeveld}, A., {Much}, R., {et~al.} 2001, \aap, 365, L36

\bibitem[{{Mewe} {et~al.}(1985){Mewe}, {Gronenschild}, \& {van den
  Oord}}]{mewe85aaps:mekal}
{Mewe}, R., {Gronenschild}, E.~H.~B.~M., \& {van den Oord}, G.~H.~J. 1985,
  \aaps, 62, 197

\bibitem[{{Nagar} {et~al.}(2005){Nagar}, {Falcke}, \&
  {Wilson}}]{4278nagar05aap}
{Nagar}, N.~M., {Falcke}, H., \& {Wilson}, A.~S. 2005, \aap, 435, 521

\bibitem[{{Nagar} {et~al.}(2002){Nagar}, {Falcke}, {Wilson}, \&
  {Ulvestad}}]{nagar02aap}
{Nagar}, N.~M., {Falcke}, H., {Wilson}, A.~S., \& {Ulvestad}, J.~S. 2002, \aap,
  392, 53

\bibitem[{{Nagar} {et~al.}(2001){Nagar}, {Wilson}, \&
  {Falcke}}]{nagar01:radobs}
{Nagar}, N.~M., {Wilson}, A.~S., \& {Falcke}, H. 2001, \apjl, 559, L87

\bibitem[{{Nandra} {et~al.}(1997){Nandra}, {George}, {Mushotzky}, {Turner}, \&
  {Yaqoob}}]{nandra97apj:variance}
{Nandra}, K., {George}, I.~M., {Mushotzky}, R.~F., {Turner}, T.~J., \&
  {Yaqoob}, T. 1997, \apj, 476, 70

\bibitem[{{Nandra} {et~al.}(2007){Nandra}, {O'Neill}, {George}, \&
  {Reeves}}]{nandra07mnras:felinesey}
{Nandra}, K., {O'Neill}, P.~M., {George}, I.~M., \& {Reeves}, J.~N. 2007,
  \mnras, 382, 194

\bibitem[{{Narayan}(2005)}]{narayan05apss:adaf}
{Narayan}, R. 2005, \apss, 300, 177

\bibitem[{{Negoro} {et~al.}(1994){Negoro}, {Miyamoto}, \&
  {Kitamoto}}]{negoro94apj:cygx1}
{Negoro}, H., {Miyamoto}, S., \& {Kitamoto}, S. 1994, \apjl, 423, L127+

\bibitem[{{Nemmen} {et~al.}(2010){Nemmen}, {Storchi-Bergmann}, {Eracleous}, \&
  {Yuan}}]{nemmen10:lineradaf}
{Nemmen}, R.~S., {Storchi-Bergmann}, T., {Eracleous}, M., \& {Yuan}, F. 2010,
  ArXiv e-prints

\bibitem[{{Nemmen} {et~al.}(2006){Nemmen}, {Storchi-Bergmann}, {Yuan},
  {Eracleous}, {Terashima}, \& {Wilson}}]{nemmen06apj:ngc1097}
{Nemmen}, R.~S., {Storchi-Bergmann}, T., {Yuan}, F., {et~al.} 2006, \apj, 643,
  652

\bibitem[{{Papadakis} {et~al.}(2008){Papadakis}, {Ioannou}, {Brinkmann}, \&
  {Xilouris}}]{papadakis08aa:sey1SED}
{Papadakis}, I.~E., {Ioannou}, Z., {Brinkmann}, W., \& {Xilouris}, E.~M. 2008,
  \aap, 490, 995

\bibitem[{{Pian} {et~al.}(2010){Pian}, {Romano}, {Maoz}, {Cucchiara}, {Pagani},
  \& {Parola}}]{pianmnras10}
{Pian}, E., {Romano}, P., {Maoz}, D., {et~al.} 2010, \mnras, 401, 677

\bibitem[{{Porquet} {et~al.}(2004){Porquet}, {Reeves}, {O'Brien}, \&
  {Brinkmann}}]{porquetaa04:21pgquasar}
{Porquet}, D., {Reeves}, J.~N., {O'Brien}, P., \& {Brinkmann}, W. 2004, \aap,
  422, 85

\bibitem[{{Ptak} {et~al.}(2004){Ptak}, {Terashima}, {Ho}, \&
  {Quataert}}]{ptak04apj:ngc3998}
{Ptak}, A., {Terashima}, Y., {Ho}, L.~C., \& {Quataert}, E. 2004, \apj, 606,
  173

\bibitem[{{Ptak} {et~al.}(1998){Ptak}, {Yaqoob}, {Mushotzky}, {Serlemitsos}, \&
  {Griffiths}}]{ptak98apj:variance}
{Ptak}, A., {Yaqoob}, T., {Mushotzky}, R., {Serlemitsos}, P., \& {Griffiths},
  R. 1998, \apjl, 501, L37+

\bibitem[{{Quataert} {et~al.}(1999){Quataert}, {Di Matteo}, {Narayan}, \&
  {Ho}}]{quataert99apj:m81ngc4579}
{Quataert}, E., {Di Matteo}, T., {Narayan}, R., \& {Ho}, L.~C. 1999, \apjl,
  525, L89

\bibitem[{{Reeves} {et~al.}(2002){Reeves}, {Wynn}, {O'Brien}, \&
  {Pounds}}]{reeves02mnras:pds456}
{Reeves}, J.~N., {Wynn}, G., {O'Brien}, P.~T., \& {Pounds}, K.~A. 2002, \mnras,
  336, L56

\bibitem[{{Sanders} {et~al.}(1989){Sanders}, {Phinney}, {Neugebauer}, {Soifer},
  \& {Matthews}}]{sanders1989ApJ:bbbseyqua}
{Sanders}, D.~B., {Phinney}, E.~S., {Neugebauer}, G., {Soifer}, B.~T., \&
  {Matthews}, K. 1989, \apj, 347, 29

\bibitem[{{Sikora} {et~al.}(2007){Sikora}, {Stawarz}, \&
  {Lasota}}]{sikora07apj:radloudagn}
{Sikora}, M., {Stawarz}, {\L}., \& {Lasota}, J. 2007, \apj, 658, 815

\bibitem[{{Steffen} {et~al.}(2006){Steffen}, {Strateva}, {Brandt}, {Alexander},
  {Koekemoer}, {Lehmer}, {Schneider}, \& {Vignali}}]{steffen06AJ:aox}
{Steffen}, A.~T., {Strateva}, I., {Brandt}, W.~N., {et~al.} 2006, \aj, 131,
  2826

\bibitem[{{Str{\" u}der} {et~al.}(2001){Str{\" u}der}, {Briel}, {Dennerl},
  {Hartmann}, {Kendziorra}, {Meidinger}, {Pfeffermann}, {Reppin}, {Aschenbach},
  {Bornemann}, {Br{\" a}uninger}, {Burkert}, {Elender}, {Freyberg}, {Haberl},
  {Hartner}, {Heuschmann}, {Hippmann}, {Kastelic}, {Kemmer}, {Kettenring},
  {Kink}, {Krause}, {M{\" u}ller}, {Oppitz}, {Pietsch}, {Popp}, {Predehl},
  {Read}, {Stephan}, {St{\" o}tter}, {Tr{\" u}mper}, {Holl}, {Kemmer},
  {Soltau}, {St{\" o}tter}, {Weber}, {Weichert}, {von Zanthier},
  {Carathanassis}, {Lutz}, {Richter}, {Solc}, {B{\" o}ttcher}, {Kuster},
  {Staubert}, {Abbey}, {Holland}, {Turner}, {Balasini}, {Bignami}, {La
  Palombara}, {Villa}, {Buttler}, {Gianini}, {Lain{\' e}}, {Lumb}, \&
  {Dhez}}]{struder01aa}
{Str{\" u}der}, L., {Briel}, U., {Dennerl}, K., {et~al.} 2001, \aap, 365, L18

\bibitem[{{Takeuchi} \& {Mineshige}(1997)}]{takeuchi97apj:adafvar}
{Takeuchi}, M. \& {Mineshige}, S. 1997, \apj, 486, 160

\bibitem[{{Terashima} {et~al.}(2000){Terashima}, {Ho}, \&
  {Ptak}}]{terashima00ApJ:liners}
{Terashima}, Y., {Ho}, L.~C., \& {Ptak}, A.~F. 2000, \apj, 539, 161

\bibitem[{{Terashima} {et~al.}(2002){Terashima}, {Iyomoto}, {Ho}, \&
  {Ptak}}]{terashima02apjs:LLAGNASCA}
{Terashima}, Y., {Iyomoto}, N., {Ho}, L.~C., \& {Ptak}, A.~F. 2002, \apjs, 139,
  1

\bibitem[{{Terashima} \& {Wilson}(2003)}]{terashima03apj:rloud}
{Terashima}, Y. \& {Wilson}, A.~S. 2003, \apj, 583, 145

\bibitem[{{Tonry} {et~al.}(2001){Tonry}, {Dressler}, {Blakeslee}, {Ajhar},
  {Fletcher}, {Luppino}, {Metzger}, \& {Moore}}]{tonry01apj:dist}
{Tonry}, J.~L., {Dressler}, A., {Blakeslee}, J.~P., {et~al.} 2001, \apj, 546,
  681

\bibitem[{{Tremaine} {et~al.}(2002){Tremaine}, {Gebhardt}, {Bender}, {Bower},
  {Dressler}, {Faber}, {Filippenko}, {Green}, {Grillmair}, {Ho}, {Kormendy},
  {Lauer}, {Magorrian}, {Pinkney}, \& {Richstone}}]{termaine02ApJ:Mbh}
{Tremaine}, S., {Gebhardt}, K., {Bender}, R., {et~al.} 2002, \apj, 574, 740

\bibitem[{{Turner} {et~al.}(2001){Turner}, {Abbey}, {Arnaud}, {Balasini},
  {Barbera}, {Belsole}, {Bennie}, {Bernard}, {Bignami}, {Boer}, {Briel},
  {Butler}, {Cara}, {Chabaud}, {Cole}, {Collura}, {Conte}, {Cros}, {Denby},
  {Dhez}, {Di Coco}, {Dowson}, {Ferrando}, {Ghizzardi}, {Gianotti}, {Goodall},
  {Gretton}, {Griffiths}, {Hainaut}, {Hochedez}, {Holland}, {Jourdain},
  {Kendziorra}, {Lagostina}, {Laine}, {La Palombara}, {Lortholary}, {Lumb},
  {Marty}, {Molendi}, {Pigot}, {Poindron}, {Pounds}, {Reeves}, {Reppin},
  {Rothenflug}, {Salvetat}, {Sauvageot}, {Schmitt}, {Sembay}, {Short},
  {Spragg}, {Stephen}, {Str{\" u}der}, {Tiengo}, {Trifoglio}, {Tr{\" u}mper},
  {Vercellone}, {Vigroux}, {Villa}, {Ward}, {Whitehead}, \&
  {Zonca}}]{turner01aa}
{Turner}, M.~J.~L., {Abbey}, A., {Arnaud}, M., {et~al.} 2001, \aap, 365, L27

\bibitem[{{Turner} {et~al.}(1999){Turner}, {George}, {Nandra}, \&
  {Turcan}}]{turner99apj:seyvar}
{Turner}, T.~J., {George}, I.~M., {Nandra}, K., \& {Turcan}, D. 1999, \apj,
  524, 667

\bibitem[{{Wang} \& {Zhang}(2003)}]{wang03MNRAS:Mbh}
{Wang}, T. \& {Zhang}, X. 2003, \mnras, 340, 793

\bibitem[{{Weisskopf} {et~al.}(2002){Weisskopf}, {Brinkman}, {Canizares},
  {Garmire}, {Murray}, \& {Van Speybroeck}}]{weisskopf02PASP}
{Weisskopf}, M.~C., {Brinkman}, B., {Canizares}, C., {et~al.} 2002, \pasp, 114,
  1

\bibitem[{{Wilms} {et~al.}(2000){Wilms}, {Allen}, \& {McCray}}]{wilms00ApJ}
{Wilms}, J., {Allen}, A., \& {McCray}, R. 2000, \apj, 542, 914

\end{thebibliography}

%
% Appendix
% 

\begin{appendix}
\section{Going through pile-up}

When  a source is  sufficiently bright  that two  or more  photons are
incident on the same CCD pixel  in a single CCD frame time, the energy
of  these photons  will be  added and  the observation  suffers photon
pile-up.  Low  energy photons will  migrate to high  energies creating
fake  high energy  events. Furthermore,  if  the added  energy of  the
photons are greater than the on-board spacecraft threshold, the photon
will be  rejected by  the spacecraft software.   Accordingly, piled-up
observations will suffer  a decrease in the total  observed count rate
and the spectral shape of the source will be distorted. Although there
is  no  simple  way  to  overcome pile-up,  two  methods  have  proved
effective.    The   statistical   method   of   \citet{davis01apjpile}
reconstruct  piled  events in  a  spectrum  by  applying a  non-linear
integral  equation  that will  connect  X-ray  source  spectra to  CCD
instrument spectra  that takes into account the  possibility of photon
pile-up. A  second more  secure approach is  to discard  spatially and
spectrally distorted piled  events from the core of  the PSF, and stay
with    non-piled    events    in     the    wings    of    the    PSF
\citep{broos98pile}.  This method  also reconstruct  temporal features
and is applicable to sources with any degree of pile-up fractions.

Pile-up  could  be  estimated  using  the  count  rate  landing  in  a
3$\times$3  pixel island  and the  CCD  readout frame  time. A  simple
converter from count rate to  pile-up fraction at the on-axis position
is   written   by   Dr.~M.~Tsujimoto   and   is  provided   as   a   C
code\footnote{\label{cr2pf}The code, \textit{cr2pf.c}, can be found at
http://www.astro.isas.jaxa.jp/$\sim$tsujimot/arfcorr.html}. Unfortunately,
we found that all of  the seven \chandra\ observations are affected by
pile-up.   To overcome  this issue,  we decided  to follow  the method
prepared    by   \citet{getman05apjsAE}\footnote{\label{pileup}   More
information            can            be           found            at
http://www.astro.psu.edu/xray/docs/TARA/ae\_users\_guide/pileup.txt.}.
The idea  is to exclude  the piled events  from the center of  the PSF
using               the               acis\_extract               (AE)
tool\footnote{http://www.astro.psu.edu/xray/docs/TARA/ae\_users\_guide.html}.

We  used  AE  tools to  sample  the  PSF  between radii  $r_{in}$  and
$r_{out}$,  the latter  fixed  to  99\% of  the  PSF, $r_{in}$  varies
between excluding 0\% of the PSF to excluding 90\% of the PSF in steps
of 5\%, whereas a  step of 1\% is used to sample  the PSF between 90\%
and  95\%. Those  annuli  were  then converted  from  PSF fraction  to
pixels. The  idea is to find  the smallest $r_{in}$  that includes the
greatest number of counts for later spectral analysis without reducing
the source flux due to pile-up.

\begin{figure}[!t]
\centerline{\includegraphics[angle=0,width=0.5\textwidth]{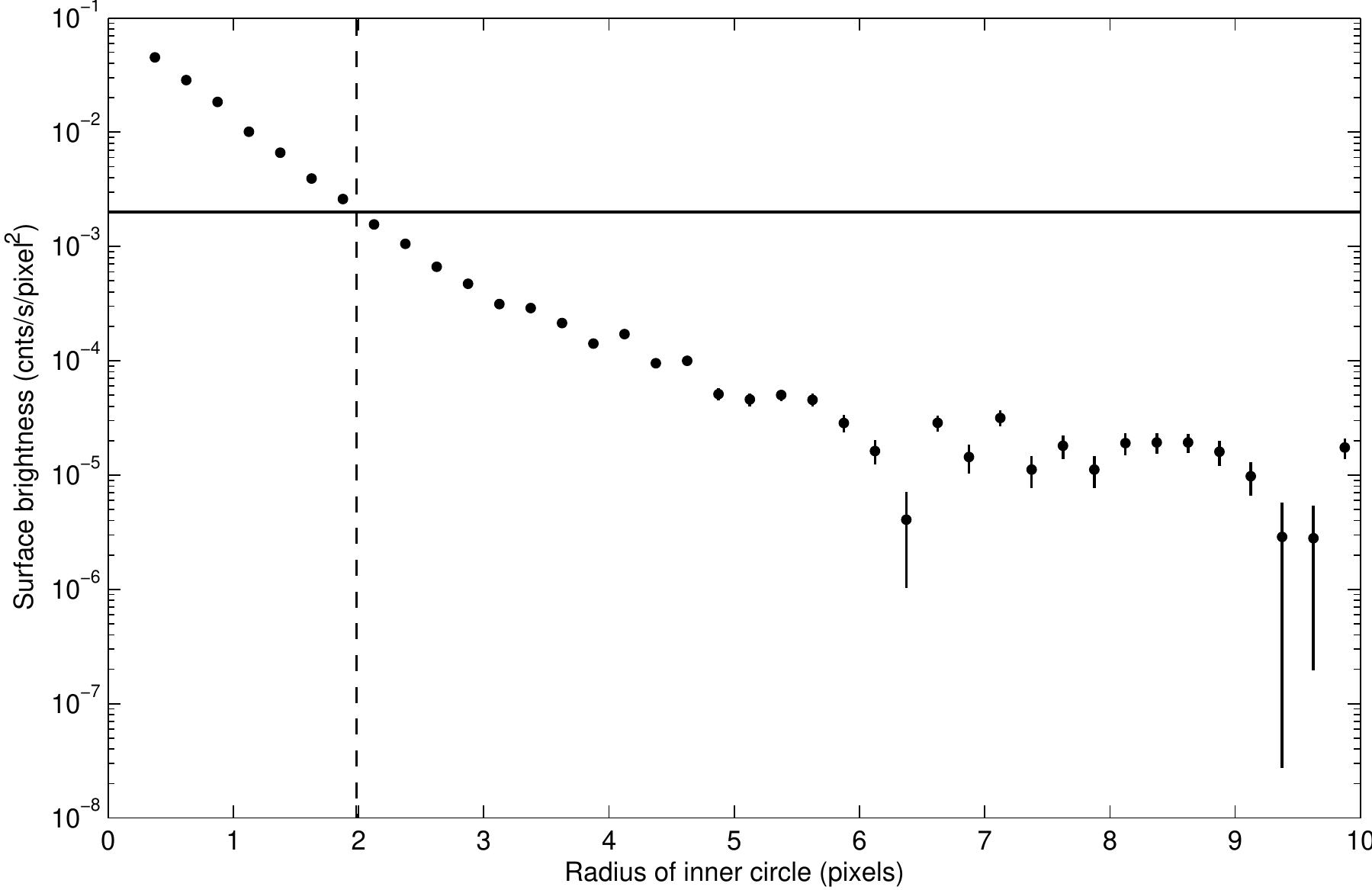}}
\caption{Surface brightness profile of NGC 4278 during the observation 7078. The solid horizontal line mark the surface brightness value that corresponds to a $\sim$2\%\ pile-up fraction. The vertical line shows the radius of the inner circle that has to be discarded.}
\label{surf-bright7078}
\end{figure}

\begin{figure}[!t]
\centerline{\includegraphics[angle=0,width=0.5\textwidth]{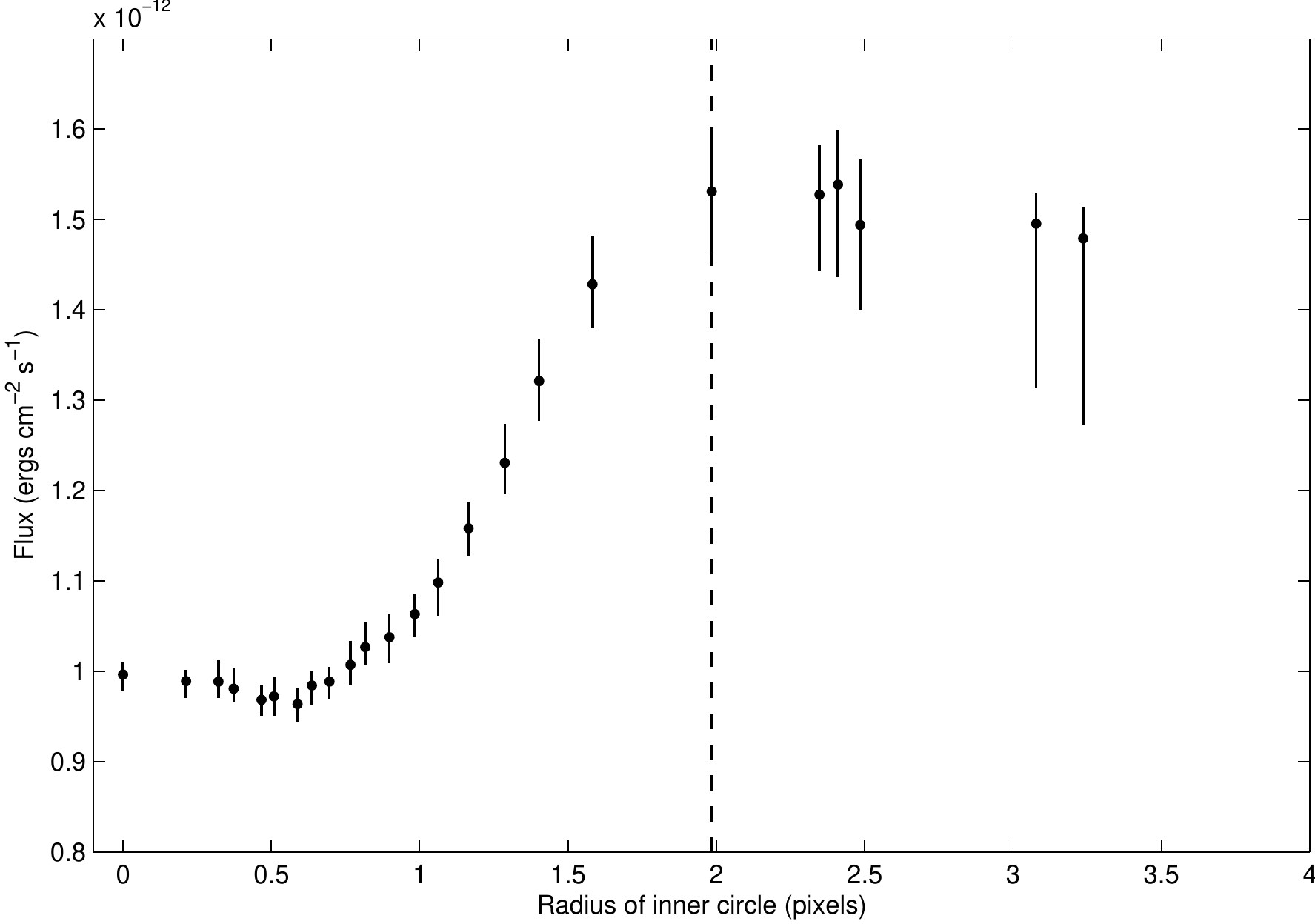}}
\caption{Inferred flux of NGC 4278 during the observation 7078 as a function of $r_{in}$ following the technique for a series of excluded core circles. The vertical solid line corresponds to the $\sim$2\%\ pile-up fraction. The flux increases to reach a plateau when the piled up core is discarded.}
\label{flux7078}
\end{figure}

Since pile-up  is related  to the surface  brightness of  the incident
photon distribution,  we created the radial profile  of annuli regions
using the \textit{dmextract} tool as described in CIAO
\footnote{\label{ciao-rad-profile} Details  on how to  obtain a radial
profile             can             be            found             at
http://cxc.harvard.edu/ciao/threads/radial\_profile/index.py.html},
and calculated  the surface brightness of every  extracted annulus. We
discarded  the   cores  giving  a  surface   brightness  greater  than
$1.2\times    10^{-3}$~counts~s$^{-1}$~pixel$^{-1}$   minimizing   the
pile-up fraction to 2\%.  Fig. \ref{surf-bright7078} shows the surface
brightness profile  of the observation 7078 with  the black horizontal
line marking the 2\%  level of pile-up.  Fig. \ref{flux7078} represents
the flux  profile of  the diminishing annuli  with the  black vertical
line marking  the 2\% level of  pile-up. We notice an  increase in the
flux when the  pile--up fraction decreases and then  reaches a plateau
after discarding the piled-up  core. Table~A.1 gives the parameters of
the \chandra\ observations corrected for pile-up.

\begin{figure}[!t]
\centerline{\includegraphics[angle=0,width=0.5\textwidth]{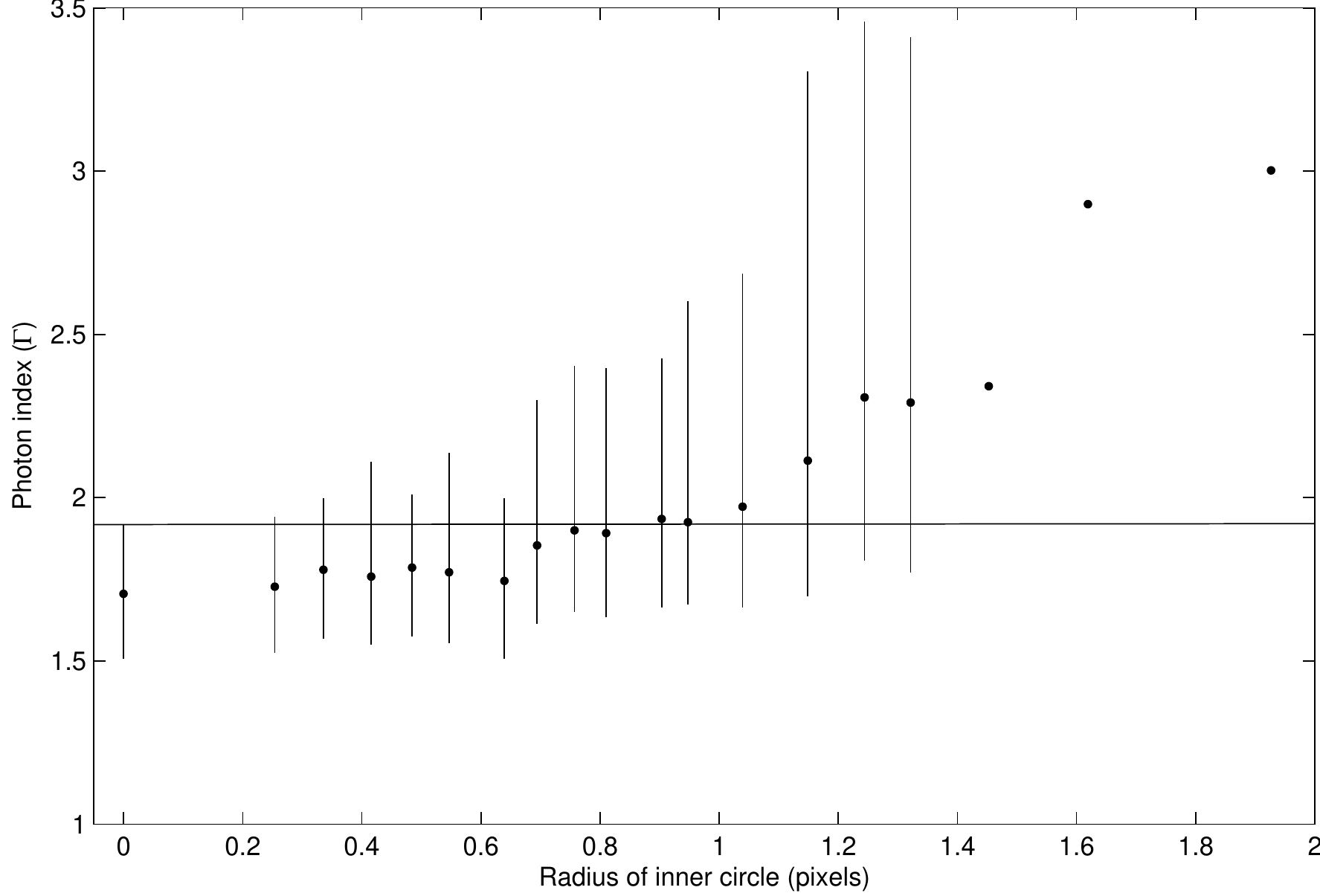}}
\caption{The photon index, $\Gamma$, as a function of $r_{in}$. Although it is clear that there is a tendency for $\Gamma$ to increase going away from the PSF center, the different values are within the error bars.}
\label{gamma398}
\end{figure}

%-----------
% Table A.1
%-----------
\begin{table*}[!th]
\label{pilup-table}
\caption{Pile-up analysis of the \chandra\ Observations.}
\newcommand\T{\rule{0pt}{2.6ex}}
\newcommand\B{\rule[-1.2ex]{0pt}{0pt}}
\begin{center}{
\begin{tabular}{c c c c c c c c}
\hline
\hline
OBS. ID \T \B & Exposure & Pile-up \% & $r_{in}$ & $r_{out}$ & Net counts & Extracted PSF fraction & Corrected count rate\\
              &  ks      &           &  Pixels  & Pixels    &            &                        & Counts~s$^{-1}$ \\
\hline
398$^a$  \T & 1.4 &11.5 & 1.45  & 2.42 & 87 & 0.2 & 0.26 \\
4741 \T & 37 &20 & 1.90  & 3.88 & 1112 &  0.1  & 0.25 \\
7077 \T & 110.5 &7.6 & 1.60  & 4.17 & 2446 & 0.15 & 0.12 \\
7078 \T & 51 &14.5 & 1.98  & 3.24 & 1272 & 0.1 & 0.19 \\
7079 \T & 105 &13.2 & 2.00  & 3.30 & 2477 & 0.1 & 0.19 \\
7080 \T & 55 &5.2 & 1.43  & 3.29 & 1202 & 0.2  & 0.09 \\
7081 \T & 110 &5.6 & 1.43  & 3.34 & 2679 & 0.2  & 0.11 \\
\hline
\end{tabular}}
\end{center}
\begin{list}{}{}
\item[Note:]$^a$The value of $r_{in}$ for this observation corresponds to a 5\% pile-up fraction.
\end{list}
\end{table*}
%-----------
% Table A.1
%-----------

Considering the small exposure time of the observation 398 that led to
a  small number  of  counts, we  decided  to discard  an inner  circle
sustaining 5\% pile-up fraction. This leaves us with 20\%\ of the PSF.
We show  in Fig.  \ref{gamma398}  that the photon index  is increasing
when we go  away from the PSF center but the  values remain within the
error  bars.   We  decided   not  to  include  this  observation  when
performing detailed spectral analysis  to avoid any misleading results
that could generate un-trustful conclusions.

\section{Surrounding medium around \src\ nucleus}

We turn now  to study the environment around the  nucleus of \src. Two
sources,  labeled source~1  and source~2  in Fig.~\ref{xmm-vs-chandra}
(hereinafter, S1 and S2),  within a 10\arcsec-radius circle around the
nucleus were  detected in the six observations  and therefore complete
X-ray spectral  study can  be possible.  We  consider the rest  of the
medium  as diffuse  emission  plus emission  from point--like  sources
detected in five  of the six observations or  less.  Source events and
spectra were  extracted, for  both S1 and  S2, from a  polygonal shape
around  the  sources  centroids  that  encircles  90\%  of  the  PSF.
Background annular regions were constructed  in the same manner as for
the  nucleus   (see  Sect.~\ref{chan-obs}).   The   spectrum  and  the
background for the diffuse emission is constructed as explained in the
AE users guide section 7.1.2.

We  fit the  six  spectra for  each of  the  two sources,  S1 and  S2,
simultaneously  with  a   power--law  model  (Fig.~B.1  and  Fig.~B.2,
respectively).   Both fit  are acceptable  with reduced  $\chi^{2}$ of
1.06 and 0.8 for 43 and 27  d.o.f for S1 and S2 respectively.  The fit
results   in   a   power--law   photon  index   of   $1.0\pm0.2$   and
$1.3_{-0.4}^{+0.7}$  for  S1  and  S2  respectively.   The  0.5--8~keV
absorption              corrected              fluxes              are
$2.6_{-0.3}^{+0.2}\times10^{-14}$~ergs~cm$^{-2}$~s$^{-1}$           and
$0.6_{-0.2}^{+0.1}\times10^{-14}$~ergs~cm$^{-2}$~s$^{-1}$  for  S1 and
S2  respectively.   Adopting  the  same  distance to  the  nucleus  of
16.7~Mpc   \citep{tonry01apj:dist},  this   results  in   a  corrected
luminosities of  $8.7\times10^{38}$ and $2\times10^{38}$~ergs~s$^{-1}$
for  S1 and  S2 respectively.   The  power--law spectral  fit and  the
luminosities derived  for these two sources make  them compatible with
the  low  mass  X--ray   binary  populations  in  elliptical  galaxies
\citep[see][for a  review]{fabbiano06aa:lmxbpop}.  The hydrogen column
density of the intrinsic absorption affecting the power--law model has
an upper limit of $\sim10^{21}$~cm$^{-2}$ for both S1 and S2.  We note
that a  disk black--body  model can  fit the spectra  as well  with an
inner disk temperature of 1.2 and 2.9~keV for S1 and S2 respectively.

The  diffuse  emission  is  fit  with a  combination  of  an  absorbed
power--law and a thermal  \textsl{mekal} component (Fig.~B.3). We find
a  reduced $\chi^{2}$  of 1.3  for 136  d.o.f.  The  power--law photon
index, $\Gamma$, is $1.8\pm0.2$ whereas the temperature of the thermal
component  is  $0.4\pm0.1$~keV.    The  discrepancy  of  this  thermal
temperature to the nucleus thermal temperature obtained with \chandra\
inside a radius of $\sim3$\arcsec\  could be the fact that the spectra
derived  for  the diffuse  emission  includes  part  of the  less--hot
interstellar medium at a larger distance from the nucleus (10\arcsec).
The intrinsic absorption derived from  the diffuse emission fit has an
upper limit of $\sim10^{21}$~cm$^{-2}$.  The corrected 0.5--8~keV flux
is $5.3\times10^{-14}$~ergs~cm$^{-2}$~s$^{-1}$ which gives a corrected
luminosity of $17.7\times10^{38}$~ergs~s$^{-1}$.

The  total  0.5--8~keV corrected  luminosity  of  the two  point--like
sources and the diffuse emission is $28.4\times10^{38}$~ergs~s$^{-1}$.
This  corresponds to  4\%\ of  the nuclear  flux during  the brightest
\chandra\ observation (obs. ID 4741)  and to 12\%\ during the faintest
one (obs. ID 7081). Now if we assume that the raise in the flux during
the  \xmm\  observation  is due  to  a  burst  in  one of  the  X--ray
point--like sources, S1  or S2, then the flux  of these objects should
increase,  in  about 8  months  period,  by a  factor  of  50 or  200,
respectively.

To check  if such an increase  in one of the  sources is qualitatively
possible we calculated the Eddington  luminosity of S1 and S2 assuming
that these sources are neutron star  LMXBs with a neutron star mass of
about   1.4~$M_{\odot}$.    This    gives   a   $L_{edd}$   of   about
1.75$\times10^{38}$~ergs~s$^{-1}$  which leads  to  a super  Eddington
ratio, $L_{X}/L_{edd}$, of 250.   Therefore, we can completely discard
the idea of an increase in the  X--ray flux of S1 and S2 that resulted
in the  high flux observed  during the \xmm\ observation.   Hence, The
raise of the  flux in the case of the  \xmm\ observation is definitely
due to  variation from the  LLAGN X--ray emission.   Consequently, S1,
S2, and the diffuse emission contribute to only 2\%\ of the total EPIC
flux  that would  explain the  less--needed thermal  component  in the
\xmm\ X-ray spectral  fit as the emission is  almost totally dominated
by the LLAGN.

\begin{figure}[!t]
\centerline{\includegraphics[angle=0,width=0.5\textwidth]{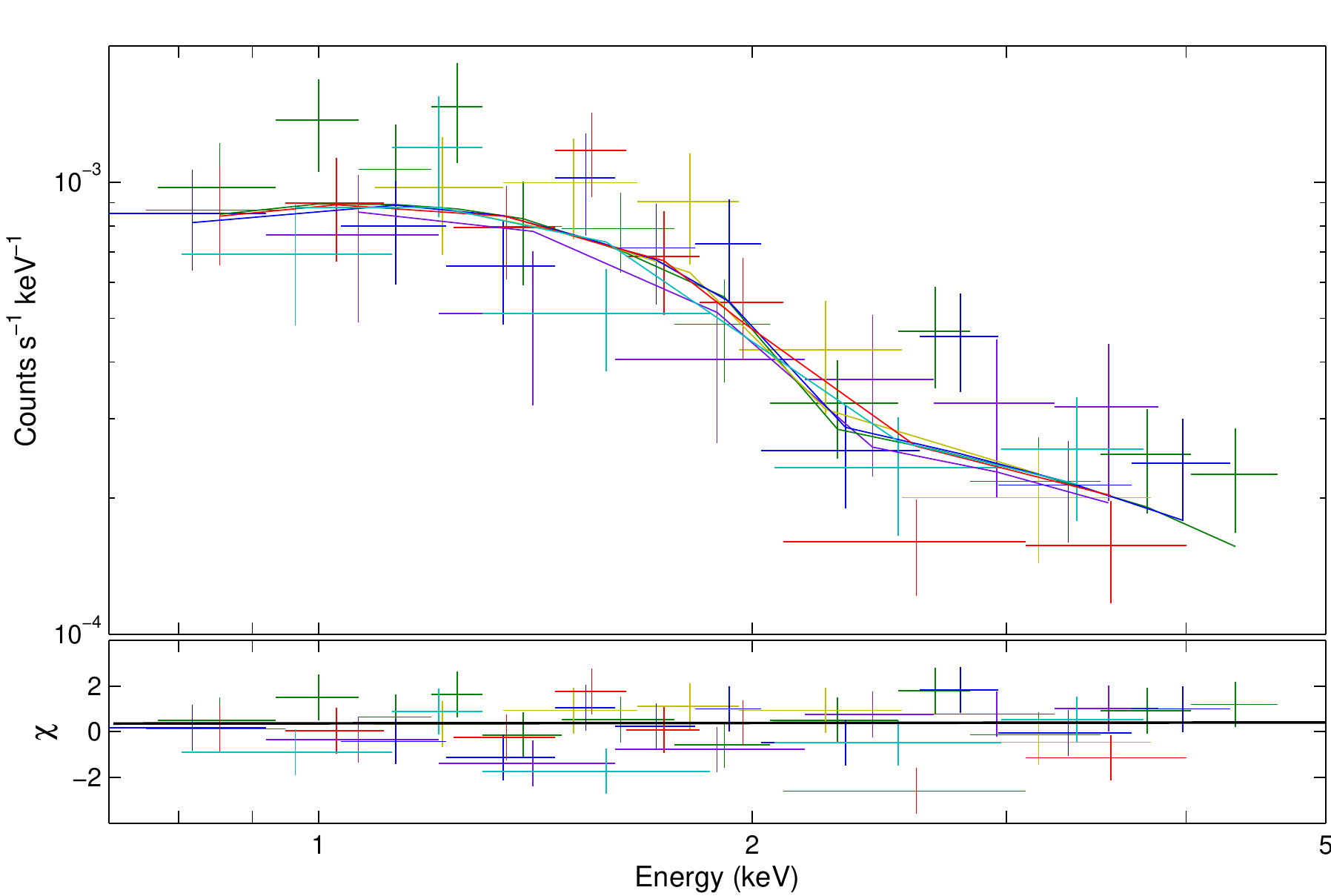}}
\caption{The best fit absorbed power--law model to S1 spectra.}
\label{s1spectra}
\end{figure}

\begin{figure}[!t]
\centerline{\includegraphics[angle=0,width=0.5\textwidth]{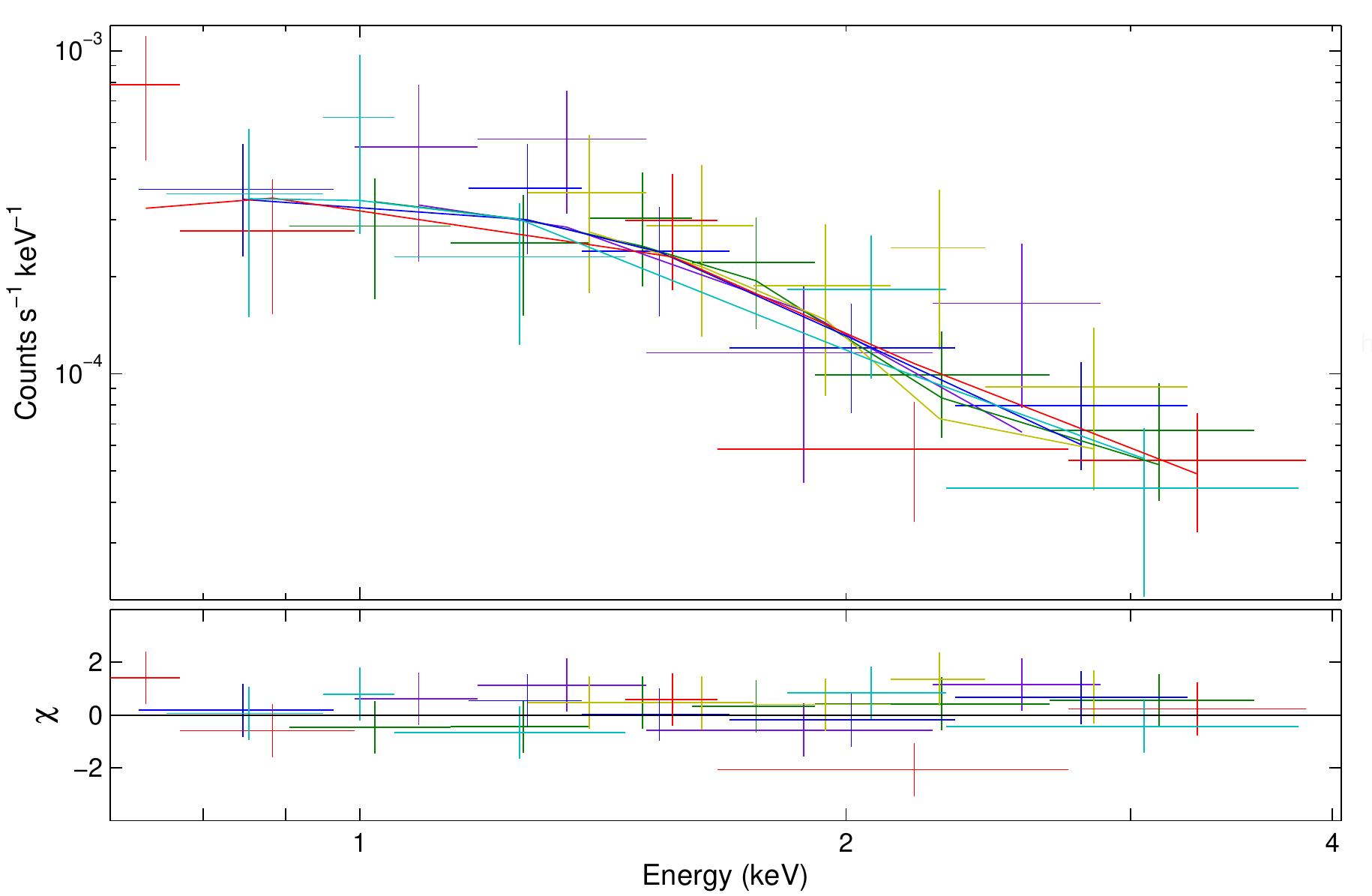}}
\caption{The best fit absorbed power--law model to S2 spectra.}
\label{s1spectra}
\end{figure}

\begin{figure}[!t]
\centerline{\includegraphics[angle=0,width=0.5\textwidth]{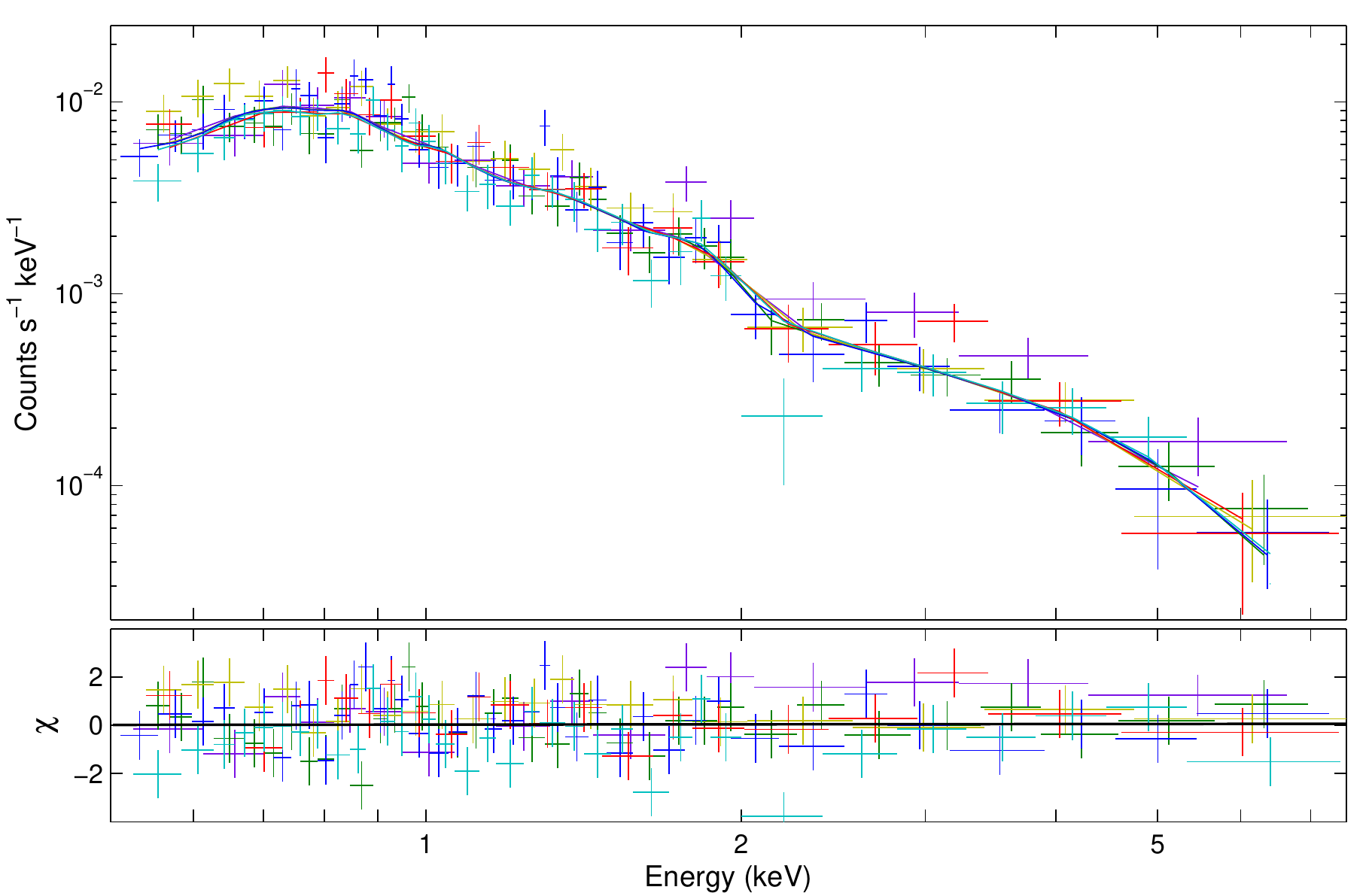}}
\caption{The best fit absorbed power--law plus a thermal component to the diffuse emission spectra.}
\label{diffspectra}
\end{figure}

\end{appendix}

\end{document}